%% file: manifold.tex
\documentclass[twocolumn]{aastex63}
\accepted{February 17, 2021}
\submitjournal{ApJ}

\usepackage[T1]{fontenc}
\usepackage[utf8]{inputenc}
\usepackage{amsmath}
\usepackage{lmodern}
\usepackage{numname}
\usepackage{xspace}

\shorttitle{Twins Embedding I}
\shortauthors{Boone et al.}
\graphicspath{{./}{figures/}}

\makeatletter
\def\input@path{{./}{latex/}}
\makeatother

\defcitealias{fakhouri15}{F15}
\defcitealias{branch06}{B06}
\defcitealias{rigault18}{R18}

\newcommand{\Bsnf}{$B_\mathrm{SNf}$\xspace}
\newcommand{\Vsnf}{$V_\mathrm{SNf}$\xspace}
\newcommand{\Rsnf}{$R_\mathrm{SNf}$\xspace}

\input{attrition_variables}
\input{peculiar_counts}
\input{sequence_peculiars}

\input{sequence_peculiars_cx}

\input{latex/explained_variance}

\begin{document}

\title{The Twins Embedding of Type Ia Supernovae I: The Diversity of Spectra at Maximum Light}

\correspondingauthor{Kyle Boone}
\email{kyboone@uw.edu}

\author[0000-0002-5828-6211]{K.~Boone}
\affiliation{Physics Division, Lawrence Berkeley National Laboratory, 1 Cyclotron Road, Berkeley, CA, 94720, USA}
\affiliation{Department of Physics, University of California Berkeley, 366 LeConte Hall MC 7300, Berkeley, CA, 94720-7300, USA}
\affiliation{DIRAC Institute, Department of Astronomy, University of Washington, 3910 15th Ave NE, Seattle, WA, 98195, USA}

\author{G.~Aldering}
\affiliation{Physics Division, Lawrence Berkeley National Laboratory, 1 Cyclotron Road, Berkeley, CA, 94720, USA}

\author[0000-0002-0389-5706]{P.~Antilogus}
\affiliation{Laboratoire de Physique Nucl\'eaire et des Hautes Energies, CNRS/IN2P3, Sorbonne Universit\'e, Universit\'e de Paris, 4 place Jussieu, 75005 Paris, France}

\author[0000-0002-9502-0965]{C.~Aragon}
\affiliation{Physics Division, Lawrence Berkeley National Laboratory, 1 Cyclotron Road, Berkeley, CA, 94720, USA}
\affiliation{College of Engineering, University of Washington 371 Loew Hall, Seattle, WA, 98195, USA}

\author{S.~Bailey}
\affiliation{Physics Division, Lawrence Berkeley National Laboratory, 1 Cyclotron Road, Berkeley, CA, 94720, USA}

\author[0000-0003-0424-8719]{C.~Baltay}
\affiliation{Department of Physics, Yale University, New Haven, CT, 06250-8121, USA}

\author{S.~Bongard}
\affiliation{Laboratoire de Physique Nucl\'eaire et des Hautes Energies, CNRS/IN2P3, Sorbonne Universit\'e, Universit\'e de Paris, 4 place Jussieu, 75005 Paris, France}

\author[0000-0002-3780-7516]{C.~Buton}
\affiliation{Univ Lyon, Univ Claude Bernard Lyon~1, CNRS, IP2I~Lyon / IN2P3, UMR~5822, F-69622, Villeurbanne, France}
    
\author[0000-0002-5317-7518]{Y.~Copin}
\affiliation{Univ Lyon, Univ Claude Bernard Lyon~1, CNRS, IP2I~Lyon / IN2P3, UMR~5822, F-69622, Villeurbanne, France}

\author[0000-0003-1861-0870]{S.~Dixon}
\affiliation{Physics Division, Lawrence Berkeley National Laboratory, 1 Cyclotron Road, Berkeley, CA, 94720, USA}
\affiliation{Department of Physics, University of California Berkeley, 366 LeConte Hall MC 7300, Berkeley, CA, 94720-7300, USA}

\author[0000-0002-7496-3796]{D.~Fouchez}
\affiliation{Aix Marseille Univ, CNRS/IN2P3, CPPM, Marseille, France}

\author[0000-0001-6728-1423]{E.~Gangler}  
\affiliation{Univ Lyon, Univ Claude Bernard Lyon~1, CNRS, IP2I~Lyon / IN2P3, UMR~5822, F-69622, Villeurbanne, France}
\affiliation{Universit\'e Clermont Auvergne, CNRS/IN2P3, Laboratoire de Physique de Clermont, F-63000 Clermont-Ferrand, France}

\author[0000-0003-1820-4696]{R.~Gupta}
\affiliation{Physics Division, Lawrence Berkeley National Laboratory, 1 Cyclotron Road, Berkeley, CA, 94720, USA}

\author[0000-0001-9200-8699]{B.~Hayden}
\affiliation{Physics Division, Lawrence Berkeley National Laboratory, 1 Cyclotron Road, Berkeley, CA, 94720, USA}
\affiliation{Space Telescope Science Institute, 3700 San Martin Drive Baltimore, MD, 21218, USA}

\author{W.~Hillebrandt}
\affiliation{Max-Planck-Institut f\"ur Astrophysik,  Karl-Schwarzschild-Str. 1, D-85748 Garching, Germany}

\author[0000-0001-6315-8743]{A.~G.~Kim}
\affiliation{Physics Division, Lawrence Berkeley National Laboratory, 1 Cyclotron Road, Berkeley, CA, 94720, USA}

\author[0000-0001-8594-8666]{M.~Kowalski}
\affiliation{Institut f\"ur Physik,  Humboldt-Universitat zu Berlin, Newtonstr. 15, 12489 Berlin, Germany}
\affiliation {DESY, D-15735 Zeuthen, Germany}

\author{D.~K\"usters}
\affiliation {Department of Physics, University of California Berkeley, 366 LeConte Hall MC 7300, Berkeley, CA, 94720-7300, USA}
\affiliation {DESY, D-15735 Zeuthen, Germany}

\author[0000-0002-8357-3984]{P.-F.~L\'eget}
\affiliation{Laboratoire de Physique Nucl\'eaire et des Hautes Energies, CNRS/IN2P3, Sorbonne Universit\'e, Universit\'e de Paris, 4 place Jussieu, 75005 Paris, France}

\author{F.~Mondon}  
\affiliation{Universit\'e Clermont Auvergne, CNRS/IN2P3, Laboratoire de Physique de Clermont, F-63000 Clermont-Ferrand, France}

\author[0000-0001-8342-6274]{J.~Nordin}
\affiliation{Physics Division, Lawrence Berkeley National Laboratory, 1 Cyclotron Road, Berkeley, CA, 94720, USA}
\affiliation{Institut f\"ur Physik,  Humboldt-Universitat zu Berlin, Newtonstr. 15, 12489 Berlin, Germany}

\author[0000-0003-4016-6067]{R.~Pain}
\affiliation{Laboratoire de Physique Nucl\'eaire et des Hautes Energies, CNRS/IN2P3, Sorbonne Universit\'e, Universit\'e de Paris, 4 place Jussieu, 75005 Paris, France}

\author{E.~Pecontal}
\affiliation{Centre de Recherche Astronomique de Lyon, Universit\'e Lyon 1, 9 Avenue Charles Andr\'e, 69561 Saint Genis Laval Cedex, France}

\author{R.~Pereira}
\affiliation{Univ Lyon, Univ Claude Bernard Lyon~1, CNRS, IP2I~Lyon / IN2P3, UMR~5822, F-69622, Villeurbanne, France}

\author[0000-0002-4436-4661]{S.~Perlmutter}
\affiliation{Physics Division, Lawrence Berkeley National Laboratory, 1 Cyclotron Road, Berkeley, CA, 94720, USA}
\affiliation{Department of Physics, University of California Berkeley, 366 LeConte Hall MC 7300, Berkeley, CA, 94720-7300, USA}

\author[0000-0002-8207-3304]{K.~A.~Ponder}
\affiliation{Department of Physics, University of California Berkeley, 366 LeConte Hall MC 7300, Berkeley, CA, 94720-7300, USA}

\author{D.~Rabinowitz}
\affiliation{Department of Physics, Yale University, New Haven, CT, 06250-8121, USA}

 \author[0000-0002-8121-2560]{M.~Rigault} 
\affiliation{Institut f\"ur Physik,  Humboldt-Universitat zu Berlin, Newtonstr. 15, 12489 Berlin, Germany}
\affiliation{Univ Lyon, Univ Claude Bernard Lyon~1, CNRS, IP2I~Lyon / IN2P3, UMR~5822, F-69622, Villeurbanne, France}

\author[0000-0001-5402-4647]{D.~Rubin}
\affiliation{Physics Division, Lawrence Berkeley National Laboratory, 1 Cyclotron Road, Berkeley, CA, 94720, USA}
\affiliation{Department of Physics, University of Hawaii, 2505 Correa Rd, Honolulu, HI, 96822, USA}

\author{K.~Runge}
\affiliation{Physics Division, Lawrence Berkeley National Laboratory, 1 Cyclotron Road, Berkeley, CA, 94720, USA}

\author[0000-0002-4094-2102]{C.~Saunders}
\affiliation{Physics Division, Lawrence Berkeley National Laboratory, 1 Cyclotron Road, Berkeley, CA, 94720, USA}
\affiliation{Department of Physics, University of California Berkeley, 366 LeConte Hall MC 7300, Berkeley, CA, 94720-7300, USA}
\affiliation{Princeton University, Department of Astrophysics, 4 Ivy Lane, Princeton, NJ, 08544, USA}
\affiliation{Sorbonne Universit\'es, Institut Lagrange de Paris (ILP), 98 bis Boulevard Arago, 75014 Paris, France}

\author[0000-0002-9093-8849]{G.~Smadja}
\affiliation{Univ Lyon, Univ Claude Bernard Lyon~1, CNRS, IP2I~Lyon / IN2P3, UMR~5822, F-69622, Villeurbanne, France}

\author{N.~Suzuki}
\affiliation{Physics Division, Lawrence Berkeley National Laboratory, 1 Cyclotron Road, Berkeley, CA, 94720, USA}
\affiliation{Kavli Institute for the Physics and Mathematics of the Universe (WPI), The University of Tokyo Institutes for Advanced Study, The University of Tokyo, 5-1-5 Kashiwanoha, Kashiwa, Chiba 277-8583, Japan}

\author{C.~Tao}
\affiliation{Tsinghua Center for Astrophysics, Tsinghua University, Beijing 100084, China}
\affiliation{Aix Marseille Univ, CNRS/IN2P3, CPPM, Marseille, France}

\author[0000-0002-4265-1958]{S.~Taubenberger}
\affiliation{Max-Planck-Institut f\"ur Astrophysik, Karl-Schwarzschild-Str. 1, D-85748 Garching, Germany}

\author{R.~C.~Thomas}
\affiliation{Physics Division, Lawrence Berkeley National Laboratory, 1 Cyclotron Road, Berkeley, CA, 94720, USA}
\affiliation{Computational Cosmology Center, Computational Research Division, Lawrence Berkeley National Laboratory, 1 Cyclotron Road MS 50B-4206, Berkeley, CA, 94720, USA}

\author{M.~Vincenzi}
\affiliation{Physics Division, Lawrence Berkeley National Laboratory, 1 Cyclotron Road, Berkeley, CA, 94720, USA}
\affiliation{Institute of Cosmology and Gravitation, University of Portsmouth, Portsmouth, PO1 3FX, UK}

\collaboration{50}{The Nearby Supernova Factory}



\begin{abstract}

We study the spectral diversity of Type~Ia supernovae (SNe~Ia) at maximum light
using high signal-to-noise spectrophotometry of \numinterpsne\ SNe~Ia from the Nearby Supernova Factory.
We decompose the diversity of these spectra into different extrinsic and 
intrinsic components, and we construct a nonlinear parametrization of the intrinsic diversity of SNe~Ia
that preserves pairings of ``twin'' SNe~Ia. We call this parametrization the ``Twins Embedding''.
Our methodology naturally handles highly nonlinear variability in spectra, such as changes in the
photosphere expansion velocity, and uses the full spectrum rather than being limited to
specific spectral line strengths, ratios or velocities.
We find that the time evolution of SNe~Ia near maximum light is remarkably similar, with 84.6\% of the variance
in common to all SNe~Ia. After correcting for brightness and color, the intrinsic variability
of SNe~Ia is mostly restricted to specific spectral lines, and we find intrinsic dispersions as low as
$\sim$0.02 mag between 6600 and 7200~\AA.
With a nonlinear three-dimensional model plus one dimension for color, we can explain
\isomapgpexpvariii\% of the intrinsic diversity in our sample of SNe~Ia, which includes several
different kinds of ``peculiar'' SNe~Ia. A linear model requires seven dimensions
to explain a comparable fraction of the intrinsic diversity.
We show how a wide range of previously-established indicators of diversity in SNe~Ia can be
recovered from the Twins Embedding.
In a companion article, we discuss how these results an be applied to standardization
of SNe~Ia for cosmology.

\end{abstract}



\section{Introduction}
\label{sec:introduction}

Type~Ia supernovae (SNe~Ia) are a relatively homogeneous class of luminous astronomical transients.
As a result of their homogeneity, SNe~Ia can be used as ``standard candles'' to infer the relative distances
to them. The use of SNe~Ia as standard candles led to the initial discovery of the accelerating
expansion of the universe \citep{riess98, perlmutter99}, and as part of a local distance ladder, SNe~Ia
provide some of the best constraints on the Hubble constant ($H_0$) \citep{riess16, riess19}.
SNe~Ia are not all identical, and understanding the diversity of SNe~Ia is crucial for our ability to
use them for cosmology. It is not currently possible to model the physics of the explosions
of SNe~Ia from first principles to the accuracy required for cosmology. Instead, cosmological analyses
of SNe~Ia rely on empirical models and corrections to parametrize the observed light curves and infer
the relative distances to individual SNe~Ia.

\subsection{The Diversity of SNe~Ia} \label{sec:intro_diversity}

Initial methods to standardize the luminosities of SNe~Ia involved correcting the observed peak brightnesses
of the SNe~Ia for correlations with the widths of their light curves \citep{phillips93} and their $B-V$ colors
at maximum light \citep{riess96, tripp98}. Current cosmological analyses fit the light curves of each SN~Ia
using an empirical model of the time-evolving spectral energy distribution (SED). The most commonly used
such model, SALT2 \citep{guy07, guy10, betoule14}, has one component $c$ for the color, and one component
$x_1$ for the intrinsic diversity that effectively captures the width of the light curve. With the SALT2
model, the luminosities of SNe~Ia can be estimated with an accuracy of $\sim$0.15~mag, and the distance to
an individual SN~Ia can be inferred with an accuracy of $\sim$8\%.

Light-curve width and color have been shown to not be sufficient to capture all of the diversity of
SNe~Ia. Most notably, the inferred SALT2 luminosities of SNe~Ia show differences of $\sim$0.1~mag when
comparing SNe~Ia in host galaxies with different masses, metallicities, colors, or star-formation
rates \citep{kelly10, sullivan10, gupta11, dandrea11, rigault13, rigault15, rigault18, childress13, hayden13, roman18}.
These observed differences with host-galaxy properties imply that there is additional unmodeled variability of
SNe~Ia.

One major open question is to determine the dimensionality of SNe~Ia, or how many different modes of
variability there are. This is a somewhat ill-posed question: SNe~Ia are the result
of highly complex explosions, and undoubtedly require a parameter space with a very large number of dimensions
to fully capture all of their intrinsic diversity.
Certain physical components presumably have a very large effect on the observed spectral timeseries of SNe~Ia,
such as the composition of the white dwarf \citep{timmes03}, total ejecta mass \citep{scalzo14},
amount of $^{56}$Ni produced \citep{arnett82},
photosphere ejecta velocity \citep{foley11b}, or asymmetry of the explosion \citep{maeda11}.
Other physical components may only be observable with highly-specialized observations with limited effect on the
full spectral timeseries, perhaps including blueshifted \ion{Na}{1} D lines \citep{phillips13}, the presence of
unburned carbon \citep{thomas11}, or polarization \citep{wang08}. For the purposes
of cosmology, it is important to consider that we are only required to model components that have a non-negligible
effect on the estimated intrinsic luminosity of SNe~Ia at maximum light. Furthermore, all of these
observational signatures of diversity are likely correlated and caused by a smaller number of underlying parameters
of the explosion.

There have been many efforts to empirically identify modes of variability of SNe~Ia
other than light-curve width and color. \citet{nugent95}
showed ratios of the equivalent widths of the \ion{Si}{2}~5972~\AA\ and \ion{Si}{2}~6355~\AA\ lines
map out a spectral sequence of SNe~Ia. \citet{branch06} used these two lines to separate
the spectra of SNe~Ia into four subgroups, and showed that at least two dimensions are required to parametrize the
intrinsic diversity of SNe~Ia. \citet{wang09} and \citet{foley11a} showed that there is diversity in the
velocity of the \ion{Si}{2}~6355~\AA\ feature that is uncorrelated with the width of the light curve and that affects
standardization. Further
studies have collected thousands of spectra of SNe~Ia and have confirmed that the intrinsic diversity of SNe~Ia
is multidimensional \citep{blondin12, silverman12, folatelli13}.

The Nearby Supernova factory \citep[SNfactory;][]{aldering02} has collected spectrophotometric
timeseries of hundreds of SNe~Ia that have enabled a wide range of new analyses of the diversity of SNe~Ia.
Using this dataset, \citet{nordin18} showed that there are at least two components of the intrinsic
diversity of SNe~Ia in the U-band that affect standardization. \citet{leget20} showed that the diversity of
13 different spectral features can be explained with an underlying three-dimensional parameter space that they call SUGAR.
\citet{saunders18} developed the SNEMO model using the full spectral timeseries from SNfactory directly,
and found that fifteen linear components are required to parametrize the diversity of this dataset.

The intrinsic diversity of SNe~Ia is complicated to
parametrize because a typical variation, such as a change in the expansion velocity of the photosphere,
leads to highly nonlinear effects in the observed spectra and photometry. Models such as SALT2, SUGAR,
or SNEMO are linear, meaning that they attempt to describe the time-evolving SED of a SN~Ia at each wavelength
and time as the sum of a set of linear components. Fitting a linear model to nonlinear phenomena will
result in a model with many more (redundant) components than
a similar nonlinear model. \citet{sasdelli16} used deep learning
to model the spectra of SNe~Ia, and showed that a nonlinear four-dimensional parameter space can capture
the intrinsic diversity of SNe~Ia as well as a 15 component linear model.
However, their analysis used the derivative of the spectrum rather than the spectrum directly
which removes information such as the brightness and color that are necessary for standardization.
Note that the linearity of a model is distinct from the linearity of standardization using
the parameters of that model. For example, \citet{rubin15} implement nonlinear standardization in terms of
SALT2 parameters, but they are still restricted to using the linear SALT2 model that cannot capture
nonlinear spectral variation.

\subsection{Supernova Twins} \label{sec:intro_sn_twins}

Using observations from SNfactory, \citet{fakhouri15} (hereafter \citetalias{fakhouri15}) introduced
an alternative method of standardizing SNe~Ia. When an underlying physical parameter
of the explosion is varied, we would expect to see a relatively smooth sequence in the spectra of SNe~Ia.
The authors developed a method of estimating the ``spectral distance'' between any pair of SNe~Ia,
and call pairs with low spectral distances ``twins''.
Standardization can then be done by inferring the luminosity of a new SN~Ia directly from
its set of twins. As long as one has a large enough reference sample to span the full range of diversity
of SNe~Ia, one can find SNe~Ia with similar spectra to any new SN~Ia. The exact functional form of how the spectrum is affected by
changes in some underlying physical parameter of the explosion is irrelevant because the twins method
only does local comparisons. The twins method does not provide a parametrization of SNe~Ia.
\citet{rubin19} showed that the statistics of the twins pairings from \citetalias{fakhouri15} are
consistent with the intrinsic variability of SNe~Ia being described by an underlying three to five dimensional
parameter space, although they did not explicitly construct this parameter space.

\subsection{Overview}

In this work, we extend the twins methodology of \citet{fakhouri15} to develop a nonlinear
parametrization of the spectral diversity of SNe~Ia. Assuming that we have a sample of observed SNe~Ia that
continuously spans the full range of intrinsic diversity of SNe~Ia, we can recover the underlying parametrization
by identifying sequences of observed spectra where each spectrum in the sequence has a small spectral distance
to its neighbors. We call this parametrization the ``Twins Embedding''.

For this analysis, we use a large dataset of high signal-to-noise spectra of SNe~Ia
from SNfactory that is described in Section~\ref{sec:manifold_dataset}. We perform a sequential
analysis to decompose the variability of SNe~Ia. First, we model
the differential time evolution of the spectra of SNe~Ia near maximum light in Section~\ref{sec:maximum_estimation},
and we estimate the spectra of all of our SNe~Ia at maximum light. In Section~\ref{sec:reading_between_the_lines},
we introduce a second procedure that we call ``Reading Between the Lines'' to estimate the contributions to
the spectra from distance uncertainties and dust extinction and produce dereddened spectra of SNe~Ia
that nominally have only intrinsic variability remaining. Finally, we perform a nonlinear decomposition of
the remaining variability of these dereddened spectra in Section~\ref{sec:decomposing_intrinsic}, and we produce
a parametrization of the intrinsic diversity of SNe~Ia that we call the ``Twins Embedding''. In Section~\ref{sec:discussion},
we explore the properties of the Twins Embedding, and show how it can be used to recover a wide range of previously-studied
indicators of intrinsic diversity of SNe~Ia. In a companion article (Boone et al. 2021; hereafter Article II), we show
how the Twins Embedding can be used to improve standardization of SNe~Ia.

\section{Dataset} \label{sec:manifold_dataset}

For this analysis, we make use of the spectrophotometric timeseries of SNe~Ia obtained by the Nearby Supernova
Factory. These spectrophotometric timeseries were collected using the Super
Nova Integral Field Spectrograph \citep[SNIFS;][]{lantz04}. The SNIFS spectroscopic channels
consist of two lenslet
integral field spectrographs \citep[IFS;][]{bacon95, bacon01}, which split a fully-filled
$6.^{\prime\prime}4 \times 6.^{\prime\prime}4$ field of view into a grid of $15 \times 15$ spatial elements.
The two channels cover the 3200--5200~\AA\ and 5100--10000~\AA\ wavelength ranges simultaneously.
A photometric channel simultaneously images the field around the IFS to monitor atmospheric transmission. The SNIFS
instrument is continuously mounted on the south bent Cassegrain port of the University of Hawaii 2.2~m
telescope on Mauna Kea.

The spectra from SNIFS were reduced using the SNfactory data reduction pipeline
\citep{bacon01, aldering06, scalzo10}. The flux calibration procedure for this pipeline is described
in \citet{buton13}, and the host-galaxy subtraction procedure is presented in \citet{bongard11}.
The spectra were corrected for Milky Way dust using the dust map from \citet{schlegel98} with an
extinction-color relation from \citet{cardelli89}.

We fit the light curve of each of the SNe~Ia in our sample using the SALT2 light curve fitter
\citep{betoule14} version 2.4 that is currently used for most cosmological analyses with SNe~Ia
\citep[e.g.,][]{scolnic18}. To perform these fits, we synthesize photometry from the spectrophotometry
in the SNfactory \Bsnf, \Vsnf, and \Rsnf bands, defined as tophat
filters with transmission for wavelengths between 4102--5100, 5200--6289, and 6289--7607~\AA\ respectively.
For this analysis, we focus specifically on spectra from the SNfactory dataset near maximum light
since \citetalias{fakhouri15} showed that twin SNe~Ia can be identified and standardized just as effectively
with a spectrum
at maximum light as with a full spectral time series. We use the SALT2 fits to determine the time of maximum light
for each SN~Ia. To ensure that we have a reasonable determination of these parameters, we require that each SN~Ia
have at least five spectra, and that the SALT2 day of maximum parameter uncertainty is less than one
restframe day. We then retain all of the spectra within five restframe days of maximum light
for our analysis.

We preprocess all of our spectra of SNe~Ia by shifting the wavelengths of the spectra to the SN~Ia's restframe,
and we adjust their brightnesses so that they appear to be at a common redshift of 0.05.
We then rebin the spectra with a common binning of 1000~km/s between 3300--8600~\AA. This results in a
total of 288 wavelength bins, and is the same binning used in \citetalias{fakhouri15} and \citet{saunders18}.
Our analysis is designed to be insensitive to the distances to SNe~Ia, so the choice of cosmological
parameters is irrelevant.

Although most of the near-maximum spectra are suitable for this study, we find that low signal-to-noise
(S/N) spectra from the SNIFS instrument can have relatively large systematic fractional systematic
uncertainties at the bluer ends of the spectra. These uncertainties appear to be primarily due to poor fits of the model in the
extraction from the CCD, which introduces a correlated offset that gets larger towards bluer wavelengths.
We find that these offsets are uncorrelated for repeated observations of the same target. 
In this analysis, we are interested in understanding the intrinsic spectral
diversity of SNe~Ia. If present, instrumental sources of spectral diversity would be recovered in such an
analysis and potentially confused with intrinsic spectral diversity. To avoid this issue, we require
that the total statistical S/N of all of the spectra used in this analysis be larger than 100 when
integrated over the bluest
500~\AA\ of the spectrum. We examined the effect of redshift, airmass, seeing, sky background level, properties
of standard stars used for calibration on a given night, photometricity of the night, moon location, detector
temperatures, time that the detector was on, and many other variables on the observed spectra. For the latest
SNfactory spectral reductions, we do not notice any significant relationships between the observed spectra and
these properties beyond S/N. One potential issue is that the requirement on S/N in the
bluest 500~\AA\ of the spectrum could bias our analysis towards a specific subtype of SNe~Ia. However, the variation
in S/N simply due to the observed brightnesses of SNe~Ia at the wide range of different redshifts
and sky brightnesses considered is much larger than the intrinsic variation in brightness in this band,
and we do not see evidence of significant selection biases when looking at e.g. SALT2 parameters.

Note that we include SNe~Ia in this analysis irrespective of whether they have been labeled as
``peculiar''. A summary of the attrition for each of these steps is
shown in Table~\ref{tab:selection_requirements}. A total of
\nummanifoldsne\ SNe~Ia pass all of the previously described selection requirements with a total of
\nummanifoldspectra\ spectra within five days of maximum light passing the S/N requirements. This dataset
is much larger than the sample of 55 SNe~Ia used in the original Twins analysis of \citetalias{fakhouri15}.

\begin{deluxetable*}{lc}
\tablecaption{Summary of sample selection requirements. The
    general selection requirements, listed in the first section of the table, are applied to all of our analyses.
    For the manifold learning analyses in Section~\ref{sec:decomposing_intrinsic},
    an additional selection requirement is imposed on the quality of the estimated spectra at maximum light.}
\label{tab:selection_requirements}
\tablehead{
    \colhead{Selection Requirement} & \colhead{Number of SNe~Ia} \\[-0.5em]
    & \colhead{Passing Requirement}
}
    
\startdata
    \input{attrition_table.tex}
\enddata
\end{deluxetable*}

\section{Estimating the Spectra of SNe~Ia at Maximum Light---The Differential Time Evolution Model} \label{sec:maximum_estimation}

Even though we have only included SNe~Ia with spectra within five days of maximum light for this analysis,
if we were to compare spectra of different SNe~Ia to each other directly, their phases could differ by as
much as ten days. The authors of \citetalias{fakhouri15} used Gaussian Process (GP) regression to
generate models of the spectral timeseries for each SN~Ia which
we could then evaluate at arbitrary phases. This method is very effective when the time series is well-sampled,
but each SN~Ia is fit completely independently of all other SNe~Ia, so the GP predictions typically
have large uncertainties when estimating the spectra of poorly-sampled time series.
As a result, the analysis of \citetalias{fakhouri15} had very strict requirements on the sampling of the SN~Ia
light curves near maximum light and was only able to use a limited subset of the SNe~Ia in the
SNfactory dataset at that time.

Instead, in this work, we build a new method of estimating the spectra of SNe~Ia at maximum light that
simultaneously models the differential time evolution of all SNe~Ia in a sample.
To build our differential time evolution model, we assume that the time evolution of the flux of SNe~Ia near
maximum light can be written as a quadratic polynomial in magnitude for each wavelength:
\begin{align}
    \label{eq:differential_evolution}
    m_i(p; \lambda_k) - m_i(0; \lambda_k) = p \cdot c_1(\lambda_k) + p^2 \cdot c_2(\lambda_k)
\end{align}
where $m_i(p; \lambda_k)$ is the spectrum of SN~Ia $i$ in magnitudes at phase $p$ and in the
wavelength bin $\lambda_k$. $c_1(\lambda_k)$ and $c_2(\lambda_k)$ are arbitrary functions of wavelength that
are the same for all SNe~Ia and that represent the time evolution of SNe~Ia near maximum light.
Note that we are modeling the differential time evolution relative to maximum light rather than the
spectra directly. Any constant multiplicative extrinsic effects such as dust reddening or
uncertainties in the distance estimate have no effect on such a differential model.

For a given SN~Ia, we label the observed flux of spectrum $s$ as $f_{\textrm{obs.},s}$. For spectra
observed with SNIFS, we find that along with typical uncorrelated measurement
uncertainties $\sigma_{\textrm{meas.},s}(p, \lambda_k)$, the individual spectra have gray dispersions $m_{\textrm{gray},s}$
in brightness with $\sigma_{\textrm{gray}}\sim0.02$~mag independent of wavelength after calibration \citep{buton13}. Furthermore, our
simple model will not be able to capture all of the diversity in spectral evolution of SNe~Ia.
To account for this, we add a term to capture the residual uncertainty of the differential time evolution model
as a function of phase. We model this uncertainty as a fraction of the observed flux
using a broken linear function $\epsilon(p; \lambda_k)$ that is fixed to zero at our reference point
of maximum light with nodes at $-5$, $-2.5$, $2.5$, and $5$ days for each wavelength bin. For computational reasons,
we implement this uncertainty as a fraction of the observed flux, but for consistency with previous literature,
we interpret it in the following text as the corresponding difference in magnitudes. Our full model
of the observed spectra and the uncertainty on them is then:

\begin{align}
    m_{\textrm{gray},s} \sim N(0; \sigma^2_{\textrm{gray}})
\end{align}
\begin{align}
    f_s(p; \lambda_k) = 10^{-0.4 (m_i(p; \lambda_k) + m_{\textrm{gray},s})}
\end{align}
\begin{align}
    \sigma_{\textrm{obs.},s}^2(p; \lambda_k) = \sigma_{\textrm{meas.},s}^2(\lambda_k) + (\epsilon(p; \lambda_k) \cdot f_s(p; \lambda_k))^2
\end{align}
\begin{align}
    f_{\textrm{obs.},s}(p; \lambda_k) \sim N(f_s(p; \lambda_k); \sigma_{\textrm{obs.},s}^2(p; \lambda_k))
\end{align}

We implement this model using the \texttt{Stan} modeling language \citep{carpenter17}, and simultaneously fit it to
our full sample of \nummanifoldspectra\ spectra within five days of maximum light for \nummanifoldsne\ different
SNfactory SNe~Ia. For each SN~Ia, we fit for
a single spectrum at maximum light $m_i(0; \lambda_k)$ that combines the information from all of the
different spectra of that SN~Ia
taken within 5 days of maximum light. With our spectra binned in 288 wavelength bins as described in
Section~\ref{sec:manifold_dataset}, this model has a total of $288 \times \nummanifoldsne$ parameters
representing the spectra at maximum light, $288 \times 2$ parameters for the $c_1(\lambda_k)$
and $c_2(\lambda_k)$ functions that
represent the time evolution of spectra, $288 \times 4$ parameters for the broken linear function that we
use to represent the model uncertainty $\epsilon(p; \lambda_k)$, \nummanifoldspectra\ parameters
for the gray offsets $m_{\textrm{gray,s}}$ of each spectrum, and finally one parameter for the gray offset dispersion
$\sigma^2_{\textrm{gray}}$.

We use \texttt{Stan} to optimize the parameters of this model to obtain the maximum \textit{a posteriori} probability (MAP)
estimate of the model parameters \citep{carpenter17}, including the predicted spectrum at the time of maximum
light of each SN~Ia. We propagate the various sources of uncertainty to obtain an estimate of the
uncertainty on the spectrum at maximum light of each SN~Ia. The recovered model parameters are shown in
Table~\ref{tab:time_evolution_parameters}.

\begin{deluxetable*}{lDDDDDDDl}
\tablecaption{Global parameters of the differential time evolution and Reading Between the Lines (RBTL) models.
A representative selection of ten lines of this table are shown here. The full table can be found in the
online version.}
\label{tab:time_evolution_parameters}
\tablehead{
    \colhead{Wavelength} & \multicolumn{12}{c}{Differential Time Evolution Model} & \multicolumn{2}{c}{RBTL Intrinsic} \\[-0.2em]
     & \multicolumn{4}{c}{Phase Evolution} & \multicolumn{8}{c}{Phase Node Uncertainties} & \multicolumn{2}{c}{Dispersion} \\[-0.2em]
    \colhead{$\lambda_k$} & \multicolumn{2}{c}{$c_1(\lambda_k)$} & \multicolumn{2}{c}{$c_2(\lambda_k)$}
    & \multicolumn{2}{c}{$\epsilon(-5.0; \lambda_k)$} & \multicolumn{2}{c}{$\epsilon(-2.5; \lambda_k)$} & \multicolumn{2}{c}{$\epsilon(+2.5; \lambda_k)$} & \multicolumn{2}{c}{$\epsilon(+5.0; \lambda_k)$} & \multicolumn{2}{c}{$\eta(\lambda_k)$} \\[-0.5em]
    \colhead{(\AA)} & \multicolumn{2}{c}{(mag)} & \multicolumn{2}{c}{(mag)} & \multicolumn{2}{c}{(mag)} & \multicolumn{2}{c}{(mag)} & \multicolumn{2}{c}{(mag)} & \multicolumn{2}{c}{(mag)} & \multicolumn{2}{c}{(mag)}
}
\decimals
\startdata
    \input{differential_time_evolution_parameters_short}
    \multicolumn{15}{c}{...} \\
\enddata
\end{deluxetable*}

\begin{figure*}
\plotone{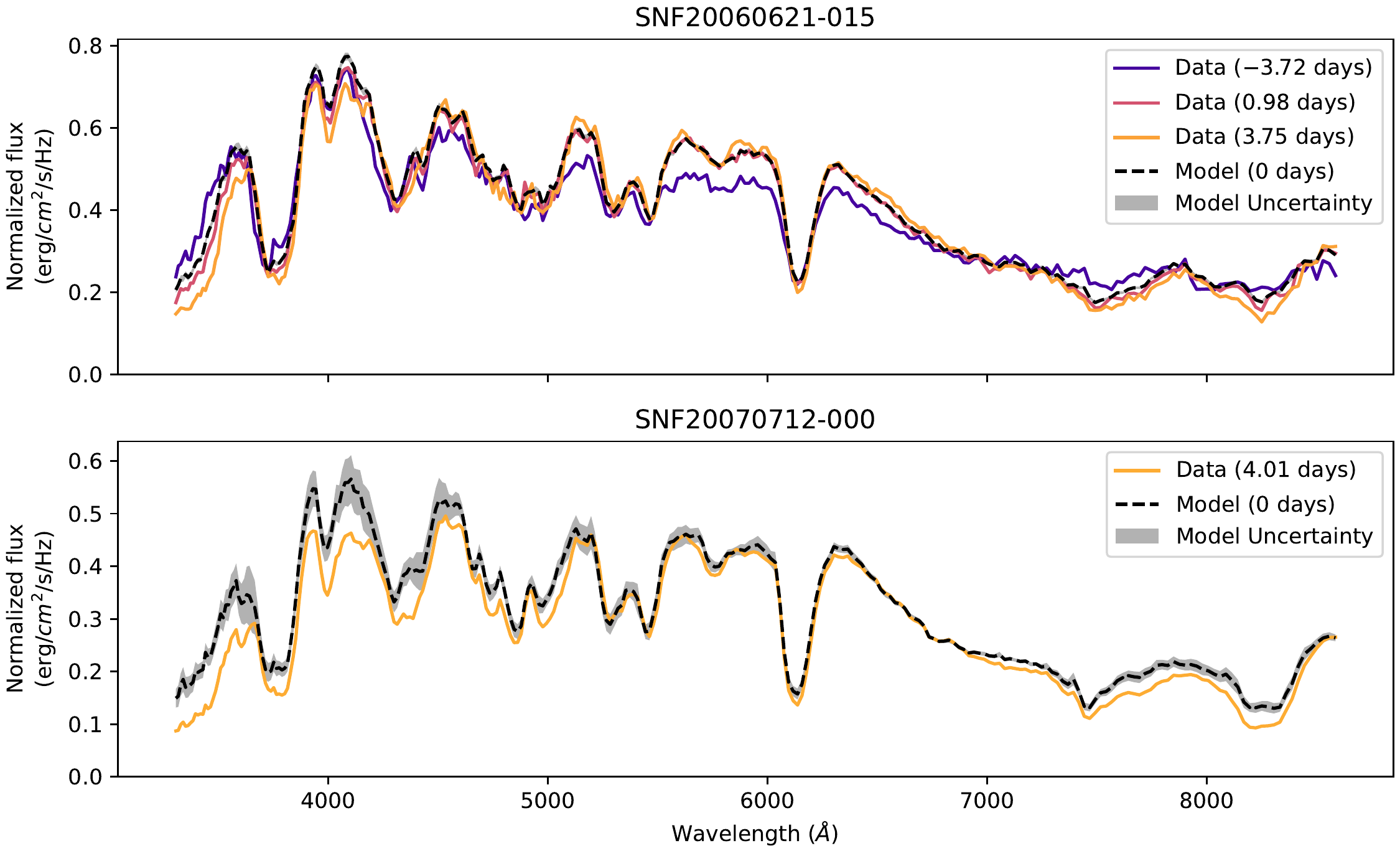}
\caption{
    Top: Estimated spectrum of SNF20060621-015 at maximum light. SNF20060621-015 has three
    different spectra passing
    the selection criteria, shown in different colors. The information from all three of these spectra is
    used to predict the spectrum at maximum light, shown with a dashed black line. A shaded gray contour
    around this dashed black line shows the uncertainty on the estimate of the spectrum at maximum light.
    Bottom: Estimated spectrum of SNF20070712-000 at maximum light. SNF20070712-000 has only
    a single spectrum passing our selection criteria, but we can still estimate the spectrum at maximum light
    using the differential time evolution model.
}
\label{fig:time_evolution_model}
\end{figure*}

This model effectively uses the SNe~Ia that have observations at multiple phases to constrain the 
$c_1(\lambda_k)$, $c_2(\lambda_k)$, and $\epsilon(p; \lambda_k)$ parameters that describe the differential
time evolution of the spectra of SNe~Ia near maximum light. If multiple spectra are available
for a given SN~Ia, then they will all be used to estimate the spectrum at maximum light. If only a single
spectrum is available for a given SN~Ia, then the spectrum will not provide any constraints on the
differential model parameters, but it can still be used to estimate the spectrum of the SN~Ia at
maximum light. Examples of this procedure are
shown in Figure~\ref{fig:time_evolution_model} for SNF20060621-015 with three observed spectra and
SNF20070712-000 with a single observed spectrum 4.77 days after maximum light.

\begin{figure*}
\plotone{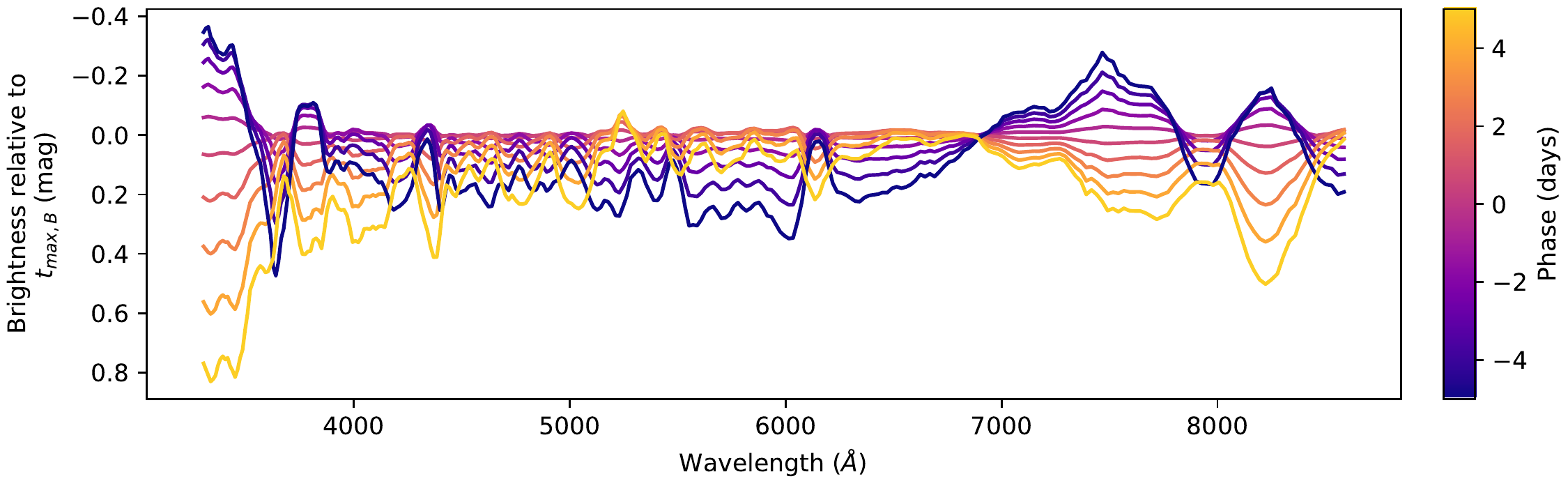}
\caption{
    Model of the differential time evolution of SNe~Ia near maximum light. The modeled
    differences are shown in different colors for phases within five days of maximum light with a spacing of one
    day. The color bar indicates which phase corresponds to which line on this plot.
}
\label{fig:spectral_evolution_model}
\end{figure*}

The recovered differential time evolution model described in Equation~\ref{eq:differential_evolution} is shown in
Figure~\ref{fig:spectral_evolution_model}. Note that we aligned all of our SNe~Ia to the SALT2-determined
time of maximum light, which is the time of maximum light in the B-band, a filter that roughly corresponds to the
integrated flux between 4000 and 5000~\AA. As expected, our model predicts that the SN~Ia gets fainter in
either direction relative to maximum light in this wavelength band. We find that the time of maximum light is
consistent from roughly 3900 to 6800~\AA. However, for wavelengths bluer than 3900~\AA\ or redder than 6800~\AA,
we find that the time of maximum light of the light curve occurs significantly earlier.
For wavelengths between 3300 to 3500~\AA, we find that the light curve declines by up to 0.2 mag/day, with this decline
becoming increasingly rapid at later phases. Similarly, in the redder bands, we find that the light curve declines by up to
0.1 mag/day in two regions of the spectrum, corresponding to the OI absorption triplet and the \ion{Ca}{2} IR triplet.
If these effects were not taken into account, then the spectrum of a SN~Ia observed 5 days after maximum light
would have systematic differences of up to 0.8 mag from the true spectrum at maximum light.

\begin{figure*}
\plotone{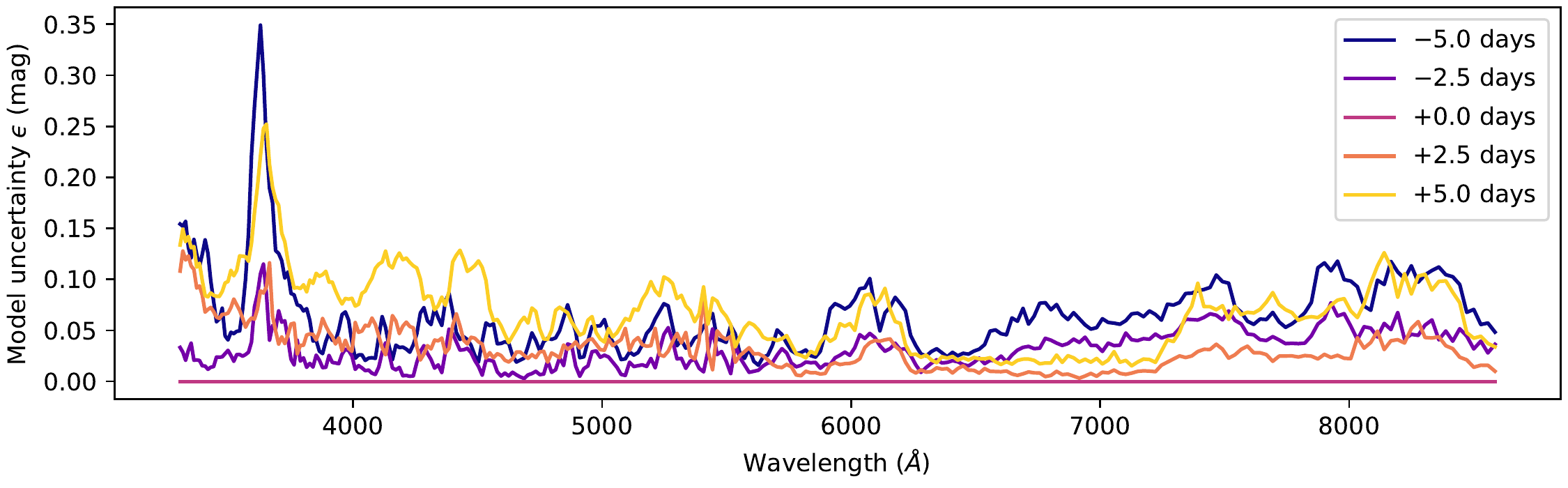}
\caption{
    Differential time evolution model uncertainties $\epsilon(p; \lambda_k)$ as a function of wavelength
    for different phases. Note that the model uncertainty at maximum light is zero by definition.
}
\label{fig:interpolation_uncertainty}
\end{figure*}

We show the uncertainty of the differential time evolution model as a function of phase in
Figure~\ref{fig:interpolation_uncertainty}. We find that our model is able to estimate the
spectra of SNe~Ia for phases within 2.5 days of maximum light with an uncertainty of less than 0.05~mag at
almost all wavelengths. For spectra 5 days away from maximum light, the model uncertainties are around 0.05~mag at most
wavelengths, but the model is unable to accurately model the time evolution of the \ion{Ca}{2} H\&K feature
around 3900~\AA. Five days before maximum light, the uncertainty of the time evolution of this feature is more than
0.35~mag for the worst wavelengths, indicating that there is significant additional variability in the time evolution
at these wavelengths that is not captured by a model that only takes into account the phase of the spectrum.
To test the accuracy of the uncertainty model, we examined pairs of spectra where one is very close to maximum light
and used the differential time evolution model to estimate the flux of a second spectrum at a different phase. We verified that
the observed residuals are consistent with the uncertainty model.

This differential time evolution model does not take the intrinsic variability of SNe~Ia into account, and
only corrects for the behavior common to all SNe~Ia. Averaged over all wavelengths, we find that 84.6\% of the variance
in the evolution of spectra near maximum light is common to all SNe~Ia and thus captured by
our differential time evolution model. As a test, we
ran a variant of our model from Equation~\ref{eq:differential_evolution} that includes a correction for SALT2 $x_1$ as
a proxy for intrinsic variability with the following functional form:

\begin{align}
    \label{eq:differential_evolution_salt}
    m_i(p; \lambda_k) - m_i(0; \lambda_k) &= p \cdot c_1(\lambda_k) + p^2 \cdot c_2(\lambda_k) \\
                                          &+ x_1 \cdot \left(p \cdot c_1(\lambda_k) + p^2 \cdot c_2(\lambda_k) \right) \nonumber
\end{align}

We find that this model explains 86.9\% of the variance in the evolution of spectra near maximum light, which
is only a slight improvement over our fiducial model. This implies that the majority of the remaining variability
is unrelated to the SALT2 $x_1$ parameter. Furthermore, we find that there are no significant differences in the results
of the rest of our analyses when we include or exclude SALT2 $x_1$ from the differential time evolution model. Note
that not including SALT2 $x_1$ or any other parametrization of the intrinsic diversity of SNe~Ia
in the differential time evolution model
slightly increases the model uncertainties, but we propagate those uncertainties to further analyses so they
should not affect our results.

\section{Reading Between the Lines} \label{sec:reading_between_the_lines}

After applying the differential time evolution model, we have estimates of the spectra at maximum light for all
of the \nummanifoldsne\ SNe~Ia in our analysis, along with uncertainties on those estimates. Our objective
is to decompose the diversity of those spectra. There are two known extrinsic contributions to the diversity of the
remaining spectra that are unrelated to the intrinsic diversity of SNe~Ia. First, in Section~\ref{sec:manifold_dataset},
we used the observed redshifts of the host galaxies of SNe~Ia
to estimate the relative distances to them and shift them to a common redshift. The observed host-galaxy redshift is not a perfect
measurement of the cosmological distance to a SN~Ia, and it also contains contributions from peculiar velocities of the
host galaxies \citep{davis11} among other factors. These uncertainties in the relative distances to SNe~Ia will introduce a change
in the overall brightness to the spectra, or a constant offset in magnitudes.

Second, interstellar dust in the SNe~Ia's host galaxies will redden the observed spectra. For the wavelengths
used in this analysis, the properties of interstellar dust can be accurately described
with a single parameter $R_V$ \citep{cardelli89}
along with a parameter $A_V$ that effectively measures the amount of dust along a line-of-sight.
\citet{chotard11} showed that the reddening of SN~Ia flux is consistent with dust with $R_V = 2.8 \pm 0.3$.
Note that differences in $R_V$ are almost entirely degenerate with an overall scale factor for the wavelengths that we
are considering in this analysis: for a fairly highly reddened SN~Ia with $E(B-V) = A_V / R_V = 0.3$, a large change in $R_V$ of 0.5
relative to a fiducial value of 2.8 introduces a nearly constant offset of $\sim$0.14~mag into the observed spectra. Differences
relative to that constant offset have a standard deviation of only $\sim$0.015~mag across different wavelengths, which
is negligible compared to the other modes of variability of SNe~Ia. Hence distance uncertainties and $R_V$ variation
have nearly degenerate effects on the optical spectra of typical SNe~Ia (a flat offset in magnitudes), and can only
be cleanly separated when the extinction is very large. For the purposes of removing extrinsic contributions to the spectra,
we can use any reasonable fiducial extinction curve. In Article~II, we model the distance uncertainties to discuss the value
of $R_V$ that best fits our sample.

In the supernova twins analysis of \citetalias{fakhouri15}, the relative difference in brightness and in dust
between each pair of SNe~Ia was measured by effectively minimizing a $\chi^2$
difference between the spectra of the two SNe~Ia while fitting for the coefficients of the difference in
brightness and dust. If two supernovae are perfect twins, then the intrinsic variability of their spectral features
should match perfectly, so only differences due to extrinsic effects such as interstellar dust extinction should remain.
Surprisingly, the estimated differences in brightness and dust between two supernovae were
consistent even when comparing two supernovae that are not twins, with differences in the estimated
brightnesses of less than 0.02~mag for even the worst pairings. This is due to the fact that the spectra of
SNe~Ia at maximum light are remarkably consistent: the spectral variability of SNe~Ia at maximum light is mostly
constrained to a handful of spectral lines, and the regions in between those lines have very little spectral
variability, as will be shown in Section~\ref{sec:rbtl_similarity}.

This result motivates a different approach to fitting for the brightness and dust of each of the supernovae in
our sample that will be discussed in this Section. Rather than compute pairwise differences
between twins, we determine a ``mean spectrum'' of a SN~Ia
at maximum light, and we compare the spectra of each supernova in our sample to this mean spectrum to determine
its brightness offset and amount of dust. To avoid our estimates of the brightness being biased by spectral features,
we simultaneously solve for the amplitude of the intrinsic dispersion of SNe~Ia at each wavelength. By weighting
by this intrinsic dispersion, we effectively deweight regions of the spectrum with large intrinsic variance,
and estimate the brightness of the spectrum and amount of dust affecting it using the regions where there is
low intrinsic diversity. We call this procedure ``Reading Between the Lines'' hereafter: RBTL). This
procedure is similar to the one developed in \citet{huang17} to compare the relative brightness and color of
SN2012cu and SN2011fe.

One caveat with this model is that any intrinsic diversity that modifies the spectrum of a SN~Ia in a way
that looks like brightness or extinction will be incorrectly labeled as extrinsic diversity at this stage. Assuming
that this intrinsic diversity also affects the spectrum in some other way, such as modifying the equivalent
widths of absorption features, we can later apply corrections to recover the intrinsic diversity that
was confused as extrinsic diversity.
An implementation of this procedure is described in Article~II. A similar procedure
is used in models like SALT2, where the true B-band maximum
brightness is determined by correcting by some function of the SALT2 $x_1$ and $c$
parameters---typically linear corrections $\alpha$ and $\beta$ for each of these parameters, see e.g. \citet{betoule14}.

\subsection{The Reading Between the Lines Model} \label{sec:rbtl_model}

The RBTL model is implemented as follows. For each supernova $i$, we
begin with a spectrum at maximum light $f_{\textrm{max.},i}(\lambda_k)$ with associated uncertainties $\sigma_{f_{\textrm{max.},i}}(\lambda_k)$
from Section~\ref{sec:maximum_estimation}.
We represent the mean spectrum of a SN~Ia at maximum light as $f_{\textrm{mean}}(\lambda_k)$. Each supernova $i$ then has a parameter
$\Delta m_i$ representing its difference in brightness compared to the mean spectrum in magnitudes, and a parameter
$\Delta \tilde{A}_{V,i}$ representing the coefficient of the extinction-color relation $C(\lambda_k)$ that best matches the supernova's spectrum
to the mean function. We choose to use the extinction-color relation $C(\lambda_k)$ from \citet{fitzpatrick99} with a
fiducial $R_V = 2.8$. The modeled flux of the spectrum at maximum light of supernova $i$ can then be written as:
\begin{align}
    f_{\textrm{model},i}(\lambda_k) = f_{\textrm{mean}}(\lambda_k) \times 10^{-0.4 (\Delta m_i + \Delta \tilde{A}_{V,i} C(\lambda_k))}
\end{align}

We assume that the intrinsic dispersion of SNe~Ia, $\eta(\lambda_k)$, is the same for all SNe~Ia and
uncorrelated in wavelength. For computational reasons, we implement this uncertainty as a fraction of the modeled flux.
However, we interpret it in the following text as the corresponding difference
in magnitudes. The total uncertainty of the spectrum at maximum light for a supernova relative to the modeled spectrum
is therefore modeled as:
\begin{align}
    \sigma^2_{\textrm{total},i}(\lambda_k) = \sigma^2_{f_{\textrm{max.},i}}(\lambda_k) + \left(\eta(\lambda_k) f_{\textrm{model},i}(\lambda_k)\right)^2
\end{align}
\begin{align}
    f_{\textrm{max.},i}(\lambda_k) \sim N(f_{\textrm{model},i}(\lambda_k); \sigma_{\textrm{total},i}^2(\lambda_k))
\end{align}

As in Section~\ref{sec:maximum_estimation}, we implement this model using the \texttt{Stan} modeling language
\citep{carpenter17}, and we use \texttt{Stan} to obtain the MAP estimate of the posterior distribution.
Finally, we apply the inverse of the magnitude and extinction corrections to obtain ``dereddened spectra''
$f_{\textrm{dered.},i}(\lambda_k)$ for each of our spectra at maximum light:
\begin{align}
    f_{\textrm{dered.},i}(\lambda_k) = f_{\textrm{max.},i}(\lambda_k) \times 10^{+0.4 (\Delta m_i + \Delta \tilde{A}_{V,i} C(\lambda_k))}
\end{align}

The resulting values of the RBTL intrinsic dispersion $\eta(\lambda_k)$ can be found in
Table~\ref{tab:time_evolution_parameters}, and the values of $\Delta m$ and $\Delta \tilde{A}_{V}$ for
each supernova can be found in Table~\ref{tab:data_table}. The dereddened spectra at maximum light 
are available on the SNfactory website at \url{https://snfactory.lbl.gov/snf/data/index.html.}

\begin{deluxetable*}{lD@{ $\pm$}DD@{ $\pm$}DD@{ $\pm$}DD@{ $\pm$}DDDD}
\tablecaption{Measurements of all of the SNe~Ia in our sample. For each SN~Ia, we show its SALT2 fit parameters,
its extracted RBTL extinction $\Delta \tilde{A}_{V}$ and magnitude residual $\Delta m$, and its coordinates in the Twins Embedding.
The uncertainties on $\Delta m$ contain contributions from both measurement uncertainties and peculiar velocities.
The first ten lines of this table are shown in this table. The full table can be found in the online version.}
\label{tab:data_table}
\tablehead{
    \colhead{} & \multicolumn{8}{c}{SALT2 Parameters} & \multicolumn{8}{c}{RBTL} & \multicolumn{6}{c}{Twins Embedding} \\[-0.5em]
    \colhead{} & \multicolumn{8}{c}{} & \multicolumn{8}{c}{Parameters} & \multicolumn{6}{c}{Coordinates} \\[-0.5em]
    \colhead{SN~Ia Name} & \multicolumn{4}{c}{$x_1$} & \multicolumn{4}{c}{$c$} & \multicolumn{4}{c}{$\Delta \tilde{A}_{V}$} & \multicolumn{4}{c}{$\Delta m$} & \multicolumn{2}{c}{$\xi_1$} & \multicolumn{2}{c}{$\xi_2$} & \multicolumn{2}{c}{$\xi_3$}
}
\decimals
\startdata
    \input{twins_manifold_coordinates_short}
    \multicolumn{23}{c}{...} \\
\enddata
\end{deluxetable*}

\subsection{Similarity of the Spectra of SNe~Ia at Maximum Light} \label{sec:rbtl_similarity}

\begin{figure*}
\plotone{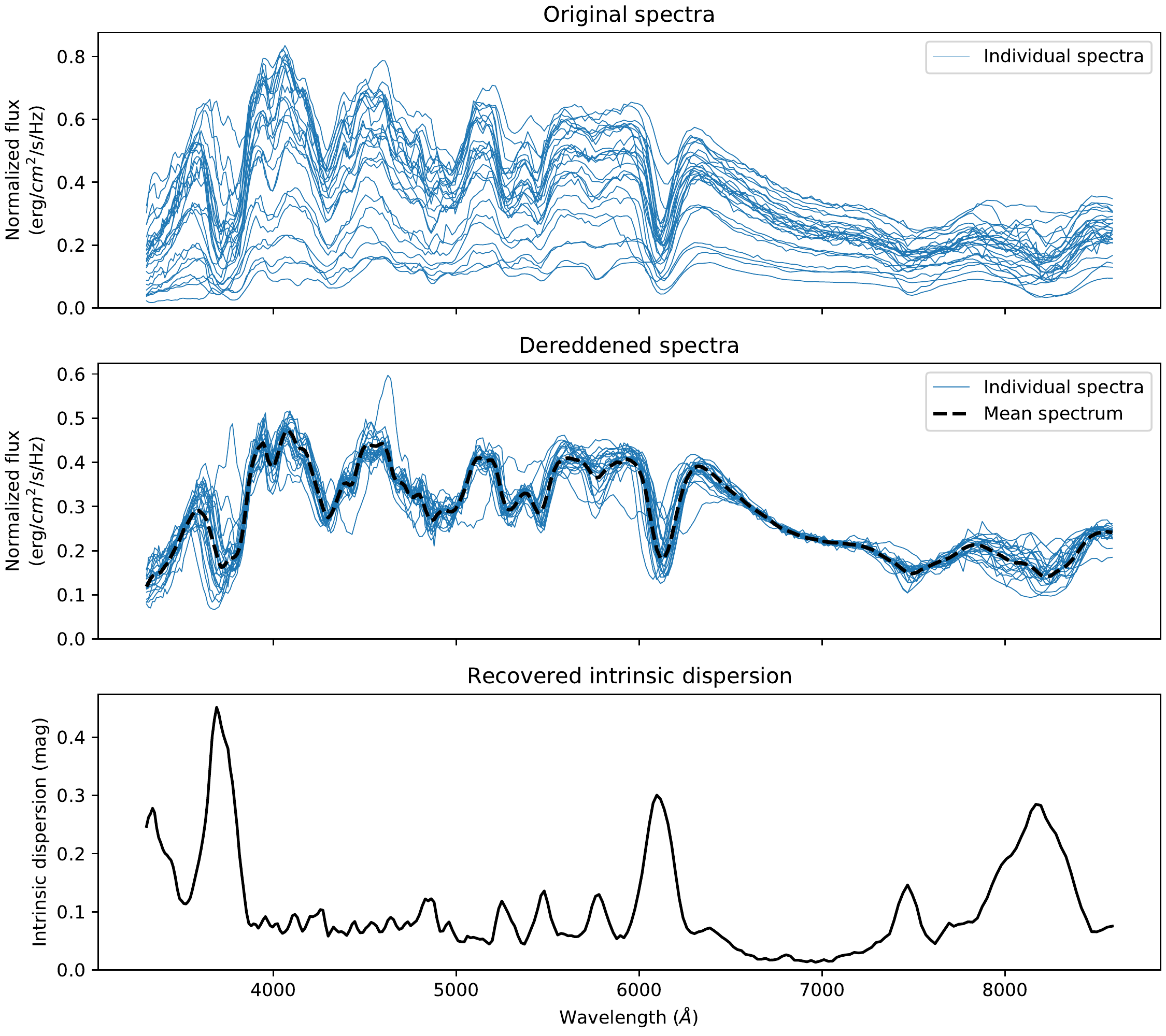}
\caption{
    Comparison of the diversity of spectra at maximum light before and after dereddening. Top:
    A sample of the original restframe spectra of SNe~Ia in our sample estimated at maximum light.
    Middle: The same
    spectra after estimating and removing the residual brightness and extinction relative to a mean spectrum
    following the RBTL procedure in Section~\ref{sec:rbtl_model}. The estimated mean spectrum is shown with a
    dashed black line. Bottom: The estimated residual intrinsic dispersion $\eta(\lambda_k)$ from the
    RBTL procedure.
}
\label{fig:dereddened_spectra}
\end{figure*}

The estimated spectra at maximum light of the \nummanifoldsne\ supernovae in our sample both before
and after dereddening are shown in Figure~\ref{fig:dereddened_spectra} along with the modeled intrinsic
dispersion. The dereddened spectra show remarkable similarity, especially in wavelength regions away
from the main absorption lines, with intrinsic dispersions of <0.10~mag at almost all wavelengths.
Note that since we fit for $\Delta m$ and $\Delta \tilde{A}_V$, any mode of the intrinsic
dispersion that affects the spectrum in a similar way to these components will not be captured
in the recovered intrinsic dispersion. The very low recovered intrinsic dispersion of
$\sim$0.02~mag between 6600 and 7200~\AA\ implies that there is almost no uncorrelated dispersion
at these wavelengths. As a result, the RBTL model effectively relies heavily on flux
measurements at these wavelengths for standardization. Perhaps not coincidentally,
in this spectral region the opacity is dominated by electron scattering rather than line absorption.
The existence of regions of low dispersion having some unique association to the physics of
radiative transfer in SNe~Ia atmospheres further supports the rationale for the RBTL technique.

The \ion{Ca}{2} H\&K lines, the \ion{Ca}{2} IR triplet and the \ion{Si}{2}
6355~\AA\ feature are the locations in the spectra of SNe~Ia with the largest intrinsic dispersions.
Interestingly, there are several regions of the spectra that are blanketed by lines but that still
show relatively low intrinsic dispersion. From 3900~\AA\ to 5900~\AA, the spectra of SNe~Ia show a variety of
absorption lines, but the intrinsic diversity is recovered to be $\sim$0.08~mag at most wavelengths,
with the exception of a handful of stronger lines that introduce diversity at up to 0.13~mag.

This model is effectively using the intrinsic diversity of SNe~Ia to
weight each wavelength when determining how to fit for the brightness and extinction of each supernova.
Hence, the lines with strong diversity are deweighted, and the model effectively ``reads between the
lines'' to estimate the brightness and extinction.

\section{A Nonlinear Model of the Intrinsic Variability of SNe~Ia} \label{sec:decomposing_intrinsic}

Using the RBTL model, we have dereddened spectra of each SN~Ia at maximum light with extrinsic contributions
from distance uncertainties and interstellar dust removed. The remaining variability between these spectra can be
interpreted as intrinsic diversity of SNe~Ia. In this section, we show that the intrinsic diversity
is inherently nonlinear, and we build a nonlinear parametric model of the intrinsic diversity of SNe~Ia.

\subsection{Manifold Learning} \label{sec:manifold}

The process of recovering a low-dimensional nonlinear parameter space from high-dimensional observations
is referred to as ``manifold learning''. Manifold learning has been shown to be very effective for constructing
nonlinear parametrizations of the spectra and light curves of astronomical objects
\citep{richards09,richards12, daniel11, matijevic12, sasdelli16}. A major challenge with nonlinear dimensionality
reduction is that any transformation of a given parametrization is an equally valid parametrization. Many
different manifold learning techniques exist, each of which imposes different assumptions and constraints on the
recovered parametrization. Our goal is not to find an ``optimal'' parametrization of SNe~Ia (which is not
well-defined), but instead one that is useful for cosmological applications.

For this analysis, we choose to construct a parametrization of SNe~Ia that
preserves the spectral distances between twin SNe as described in \citetalias{fakhouri15}. We
define the spectral distance $\gamma_{ij}$ between two SNe~Ia labeled $i$ and $j$ as:
\begin{align}
    \label{eq:spectral_distance}
    \gamma_{ij} = \sqrt{\sum_k \left(\frac{f_{\textrm{dered.},i}(\lambda_k) - f_{\textrm{dered.},j}(\lambda_k)}{f_{\textrm{mean}}(\lambda_k)}\right)^2}
\end{align}
Our goal is to construct a parametrization of SNe~Ia where the Euclidean distance between the
coordinates of any two twin SNe~Ia is equal to the spectral distance between those two SNe~Ia.
We do not require that the spectral distances of non-twins be preserved.

To accomplish these goals, we make use the Isomap algorithm from \citet{tenenbaum00}.
This algorithm is designed to embed points from a high-dimensional space into a
low-dimensional one while preserving the distances between nearby points in the high-dimensional space.
By using the spectral distance of Equation~\ref{eq:spectral_distance} as the distance measure
for the Isomap algorithm, we will therefore generate an embedding that preserves the distances between twin SNe~Ia.

The Isomap algorithm proceeds as follows. First, for each SN~Ia, we find its $K$ nearest neighbors using
spectral distances. For each pair of SNe~Ia in the sample, we then find
the shortest path between the two SNe~Ia passing through pairs of neighbors. We calculate the ``geodesic distance''
for each pair of SNe~Ia as the sum of the spectral distances between neighbors along each step in this path.
After computing the geodesic distances between each pair of SNe~Ia, we obtain a distance matrix. To generate an
embedding, we center the distance matrix and compute its eigendecomposition (see \citet{tenenbaum00} for details).
For a $D$-dimensional embedding, we keep the $D$ eigenvectors of the centered distance matrix with the largest
eigenvalues. Each eigenvector then corresponds to a ``component'' of the embedding and captures a distinct
mode of variability of SNe~Ia. The values of the eigenvectors are the ``coordinates'' of each SN~Ia within the
embedding.

The Isomap algorithm has two parameters that must be set to produce an embedding:
the number of neighbors $K$, and the dimensionality $D$ of the embedding.
As will be discussed in Section~\ref{sec:stability}, the number of neighbors
$K$ does not have a major impact on the resulting
embedding, and for a range of different values tested between $\sim$6 and 50 we obtain nearly
identical embeddings. We choose to use 10 neighbors for the rest of our analysis.
We will discuss the dimensionality of SNe~Ia in Section~\ref{sec:num_components}.

The Isomap algorithm generates a low-dimensional embedding of SNe~Ia, but it does not produce a model of the spectrum
of an SN~Ia given its coordinates in the embedding. Instead, to do this we use Gaussian Process (GP) regression to model the spectra
as a function of the Isomap coordinates. The details of this model are described in Appendix~\ref{sec:gaussianprocess}.
We use a separate GP model for each wavelength and we optimize the hyperparameters of each GP independently.

\subsection{Sample for Manifold Learning Analyses} \label{sec:isomap_sample}

For some of the supernovae in our sample, the uncertainty on the spectrum at maximum light is comparable to the
recovered intrinsic dispersion $\eta(\lambda_k)$. The Isomap algorithm is unable to take this
uncertainty into account, so if spectra with large uncertainties are included in our analysis,
we could confuse
intrinsic diversity with the uncertainty in our estimate of the spectrum at maximum light. To mitigate this, we
remove any supernovae at this stage of the analysis whose uncertainties on the spectrum at maximum light
are large compared to the intrinsic dispersion. We choose to require that the total measurement variance
of the estimate of the spectrum at maximum light be less than 10\% of the total intrinsic variance of SNe~Ia.
Choosing this fractional variance threshold
requires somewhat of a trade off: with better measured spectra, we could potentially recover more components of the
intrinsic dispersion, but a stricter threshold reduces the number of SNe~Ia in the sample and thus our ability
to reconstruct the parameter space. Out of the original sample
of \nummanifoldsne~SNe~Ia, \numinterpsne~SNe~Ia have an uncertainty
on their spectrum at maximum light that passes this stringent fractional variance threshold. As
will be shown in Section~\ref{sec:stability}, whether this cut is applied has almost no effect on
the embedding coordinates for the SNe~Ia that pass it.

Note that in the RBTL analysis we included the estimates of the uncertainties for the spectra at maximum light, so
the spectra that were cut due to the fractional variance threshold will not have a major impact on the RBTL analysis.
We retrained the RBTL model on only the spectra that pass the uncertainty of the spectrum at maximum light requirement, and
found that the changes were negligible (the estimated brightnesses change by <0.005~mag). As a result, we choose to use
the RBTL model trained on all of the spectra for further analysis.

\subsection{Dimensionality of the SN~Ia Population at Maximum Light} \label{sec:num_components}

\begin{figure}
\plotone{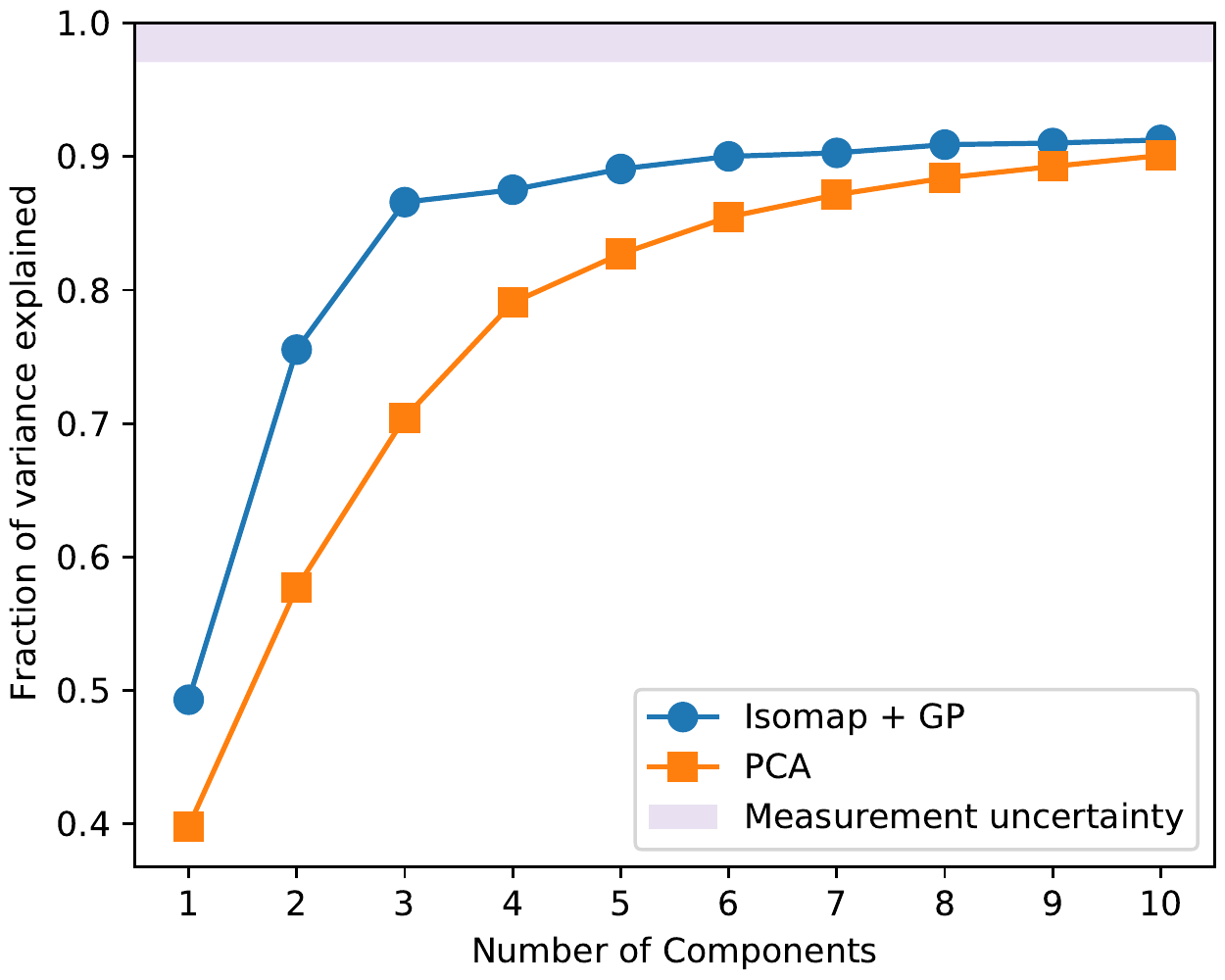}
\caption{
    Fraction of the intrinsic variance of SNe~Ia explained by different models. We show the results
    for both the nonlinear Isomap + GP model and a linear PCA-based model. The shaded area at the top
    of the plot corresponds to the fraction of variance explained by measurement noise.
}
\label{fig:explained_variance_comparison}
\end{figure}

We use the Isomap + GP model to investigate the dimensionality of SNe~Ia.
For a given number of Isomap components $D$, we examined what fraction of the variance is explained by the
model. The results of this procedure are shown in Figure~\ref{fig:explained_variance_comparison}.
We find that the first three components of the Isomap + GP model each explain a significant fraction
(\isomapgpexpvarindivi, \isomapgpexpvarindivii, and\ \isomapgpexpvarindiviii\% respectively)
of the total variance, and together they explain
\isomapgpexpvarfulliii\% of the total variance. Measurement uncertainty accounts for \isomapgpnoise\% of the remaining
variance for this sample, so our model explains \isomapgpexpvariii\% of the intrinsic variance when this
is taken into account. Additional components beyond the third do not explain a significant amount of the remaining variance.

For comparison purposes, we also perform a linear PCA decomposition of the same data. We find that the linear
model requires seven components to explain as much of the variance as our three component nonlinear model, which
supports our claim that there is a significant amount of nonlinear variability in the spectra of SNe~Ia.

\begin{figure*}
\plotone{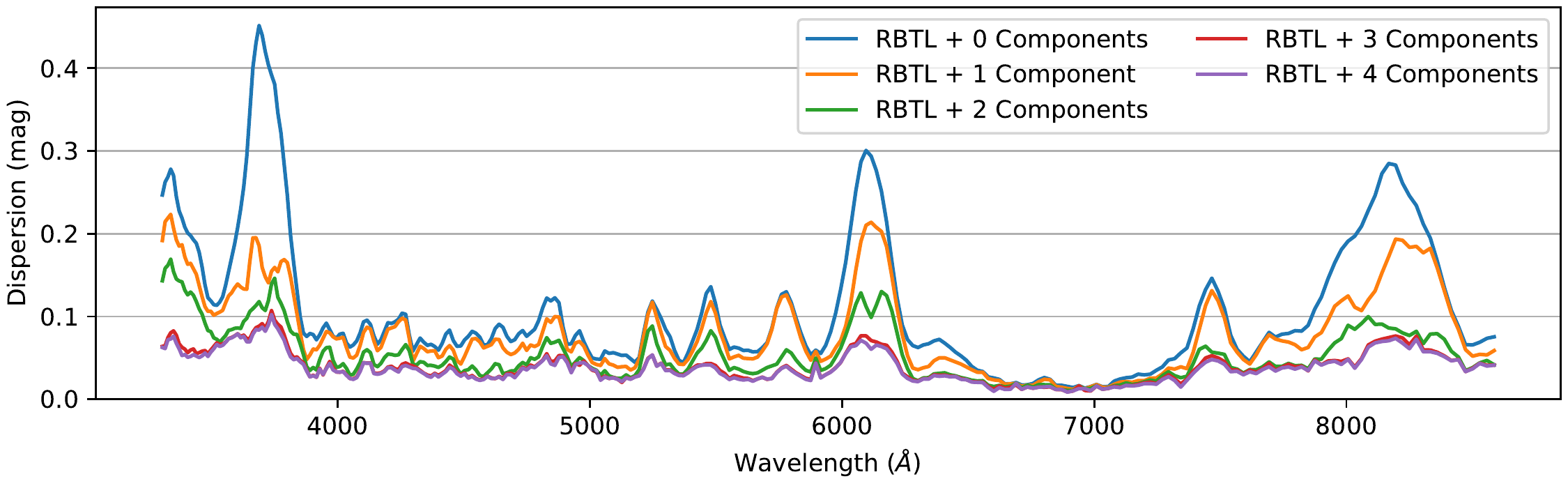}
\caption{
    Residual intrinsic dispersion in the spectra of SNe~Ia for Isomap + GP models with different
    numbers of components.
}
\label{fig:isomap_gp_dispersions}
\end{figure*}

We show the unexplained dispersion as a function of wavelength in Figure~\ref{fig:isomap_gp_dispersions}.
After the RBTL procedure, the residual dispersion is $\sim$0.3 mag for the \ion{Ca}{2} and \ion{Si}{2} features.
In contrast, with the three component model Isomap + GP model, there is less than $0.1$~mag of residual dispersion at the
wavelengths associated with these features, and the dispersion is less than $0.05$ mag at almost all wavelengths. Again,
adding components beyond the third has a negligible effect on the intrinsic dispersion.

Alternatively, we look at how well the spectral distances from Equation~\ref{eq:spectral_distance}
are preserved in the embedding. We find that the Euclidean distances between two SNe~Ia in the Isomap
embedding have Pearson correlations with the \citetalias{fakhouri15} spectral distances of 0.77 for 1 component,
0.90 for 2 components and 0.95 for either 3 or 4 components with very little improvement if additional components are
added. With at least three components, we are therefore able to accurately preserve the spectral distances
(and thus twin pairings) of \citetalias{fakhouri15}.

\citet{sasdelli16} previously showed that deep learning can produce a nonlinear four-dimensional
representation of SNe~Ia. However, their analysis modeled the derivatives of the
spectra in wavelength rather than modeling the spectra directly, and didn't include an explicit model
of how spectra vary near maximum light or a means of measuring the brightness and color of a spectrum
(which is necessary for cosmological applications).
Because we explicitly model the extrinsic contributions to the spectrum, our model is able to explain
significantly more of the variance
in the spectra: with three components we can explain \isomapgpexpvarfulliii\% of the intrinsic variance compared to
82\% with four components for the analysis of \citet{sasdelli16}. Furthermore, the distances between two SNe~Ia
in the Twins Embedding capture the spectral distances of \citetalias{fakhouri15} while they have no meaning
in the analysis of \citet{sasdelli16}.

\subsection{Reconstructing Spectra}

\begin{figure*}
\epsscale{1.15}
\plotone{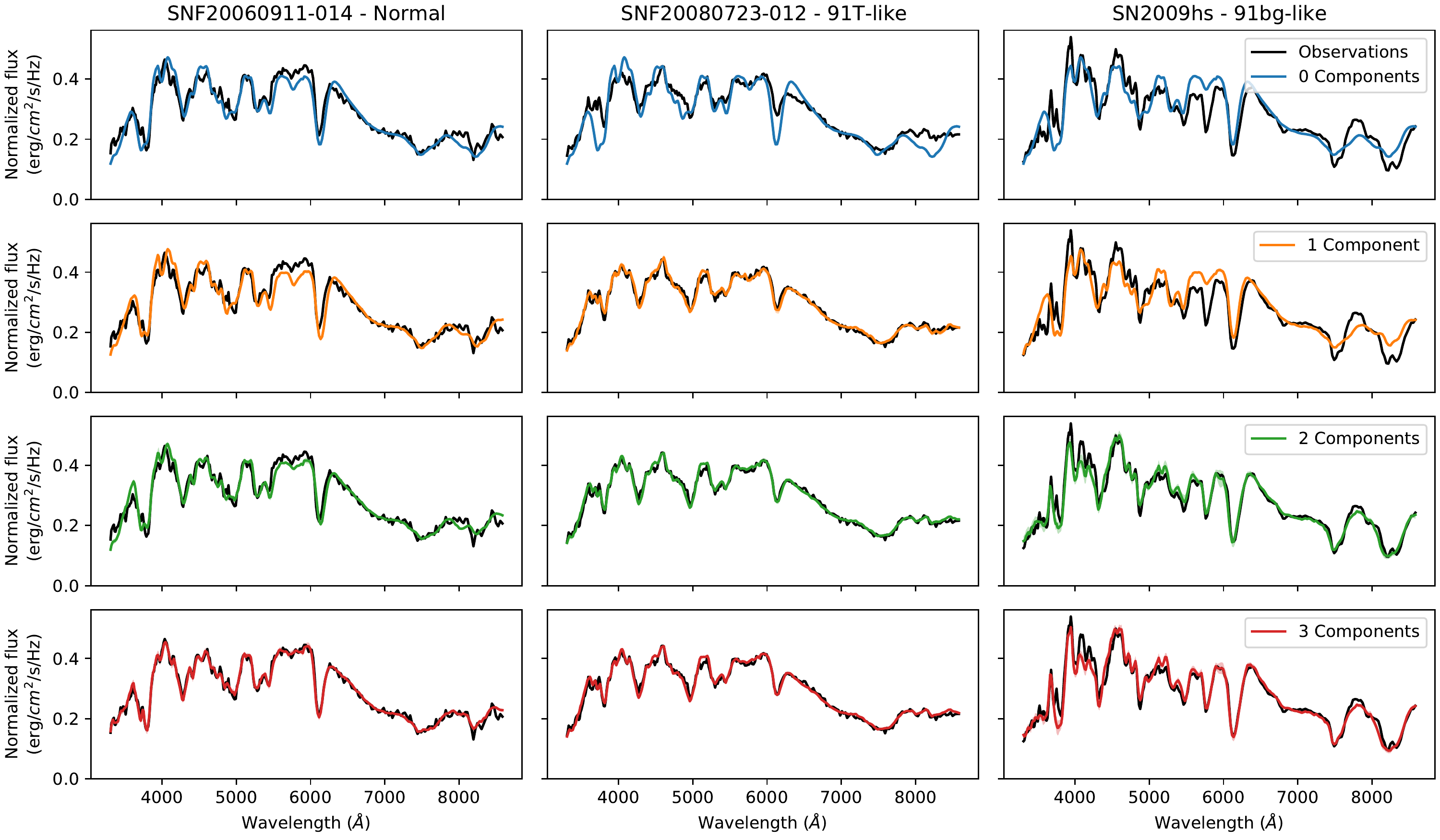}
\caption{
    Examples of the leave-one-out Isomap + GP model predictions for SNe~Ia with different numbers of
    components. The original
    observations are shown in black, while the model is shown in color. Each column corresponds to a different
    SN~Ia. The first row shows the results of applying the RBTL model to capture brightness and color.
    Subsequent rows show the Isomap + GP model for different numbers of components.
}
\label{fig:reconstructed_spectra}
\end{figure*}

We can use the GPs to predict the spectrum of an SN~Ia given its coordinates from the Isomap
decomposition. To test how well our model would perform on new observations, we generate leave-one-out
predictions where we condition the GP on spectra from all of the SNe~Ia in our sample except for one SN~Ia,
and predict the spectrum at the Isomap coordinates of the remaining SN~Ia. Results of this procedure
for three SNe~Ia are shown in Figure~\ref{fig:reconstructed_spectra}. In this plot, we show the results for
a ``normal'' SN~Ia, a 91T-like SN~Ia \citep{filippenko92b} and a 91bg-like SN~Ia \citep{filippenko92a}. We
find that the three-component model is able to predict accurate spectra for the full range
of SNe~Ia including 91T-like and 91bg-like SNe~Ia.

\begin{figure*}
\epsscale{1.15}
\plottwo{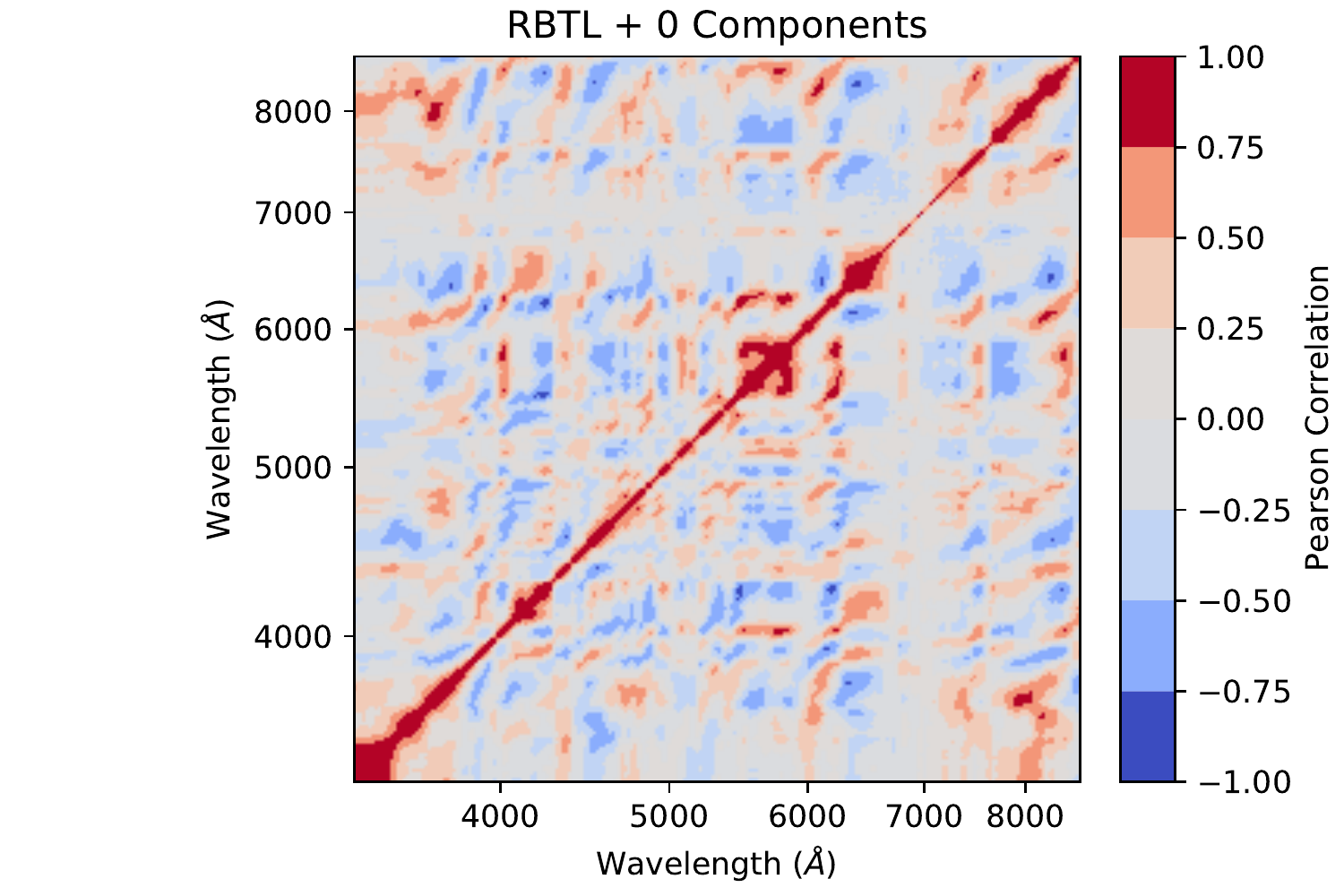}{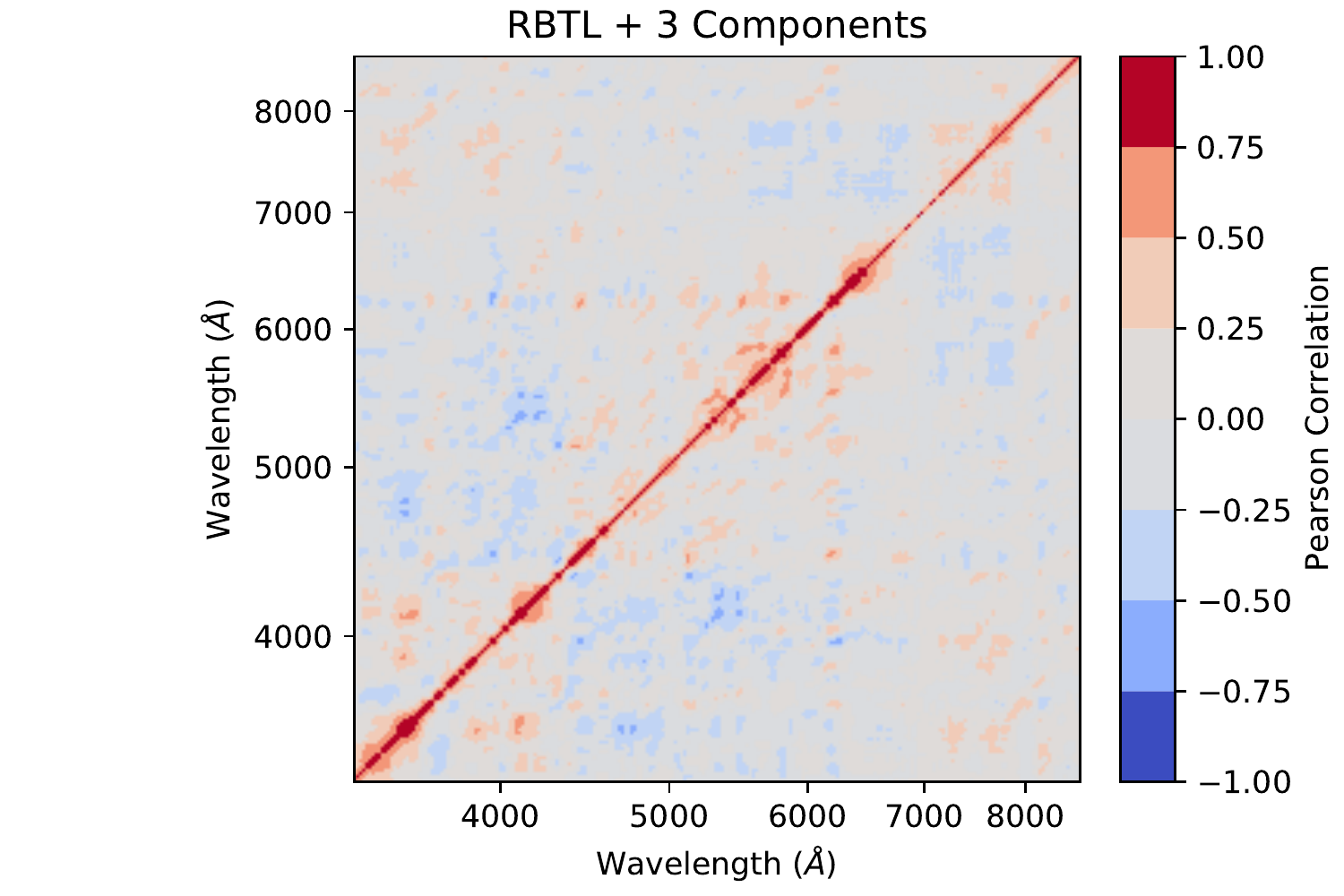}
\caption{
    Correlation matrix of the residuals of spectra of SNe~Ia with different models.
    Left panel: Correlation matrix for the RBTL model. Right panel: Correlation matrix for the three component
    Isomap + GP model. We find that a three component Isomap + GP model is able to capture the vast
    majority of the variance that is correlated across different wavelengths.
}
\label{fig:correlation_gp}
\end{figure*}

There is still a small amount of residual dispersion after applying any of these models.
To study this, we performed leave-one-out predictions for all of the SNe~Ia in our sample, and evaluated
the correlation matrix of the residuals. The results of this procedure are shown in
Figure~\ref{fig:correlation_gp}. For the base RBTL model (left panel of Figure~\ref{fig:correlation_gp}),
we see strong off-diagonal structure in the
correlation matrix implying that the residuals at different wavelengths are highly correlated. In contrast,
for the three component Isomap + GP model (right panel of Figure~\ref{fig:correlation_gp})
there is very little correlation between different wavelengths.
This implies that the remaining variance is mostly uncorrelated across wavelengths,
and explains why adding additional components does little to improve the model.

We conclude that a three-dimensional embedding is sufficient to explain the vast majority of intrinsic diversity of SNe~Ia
at maximum light (along with extrinsic contributions from dust and brightness removed in
Section~\ref{sec:reading_between_the_lines}). As this embedding was effectively constructed using pairs of
twin SNe~Ia, we refer to it hereafter as the ``Twins Embedding''. We label the three components
of the Twins Embedding $\xi_1$, $\xi_2$, and $\xi_3$. The RBTL parameters and coordinates
of each SN~Ia in the Twins Embedding can be found in Table~\ref{tab:data_table}.

\subsection{Stability of the Model} \label{sec:stability}

\begin{figure}
\plotone{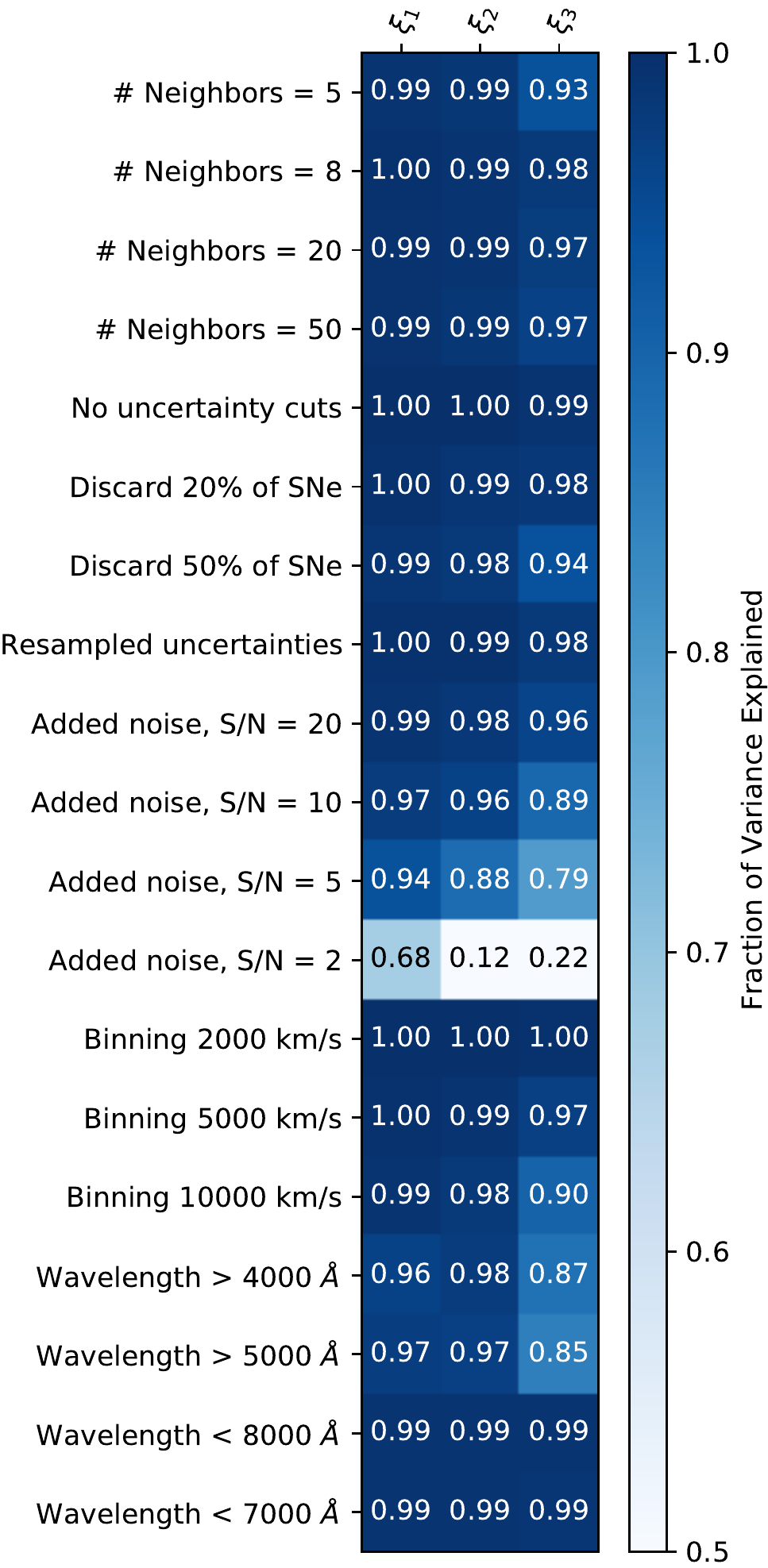}
\caption{
    Comparison of how various modifications of the dataset affect the Twins Embedding.
    For each modification, labeled on the left, we generate a three-component embedding and calculate the
    fraction of variance of the original Twins Embedding that can be explained by a transformation
    of the alternative embedding (see text for details).
}
\label{fig:cross_validation_manifold}
\end{figure}

As described in Section~\ref{sec:manifold}, nonlinear dimensionality reduction is challenging
because there is in general no unique solution. We investigated how stable the Twins Embedding is by making
various modifications to the input dataset and applying our algorithms to generate variants on the embedding.
In general, there is no guarantee
that the recovered embedding will be axis-aligned with our original one, especially if we modify the
binning of the spectrum which will change the weights of different regions. To perform a robust quantitative
comparison, we fit a GP model to predict the coordinates in the original embedding from the coordinates
in some alternative embedding. We use the same GP model that was used to predict the flux of each
spectrum across the Twins Embedding, the details of which can be found in Appendix~\ref{sec:gaussianprocess}.
For each variant, we calculate the fraction of variance that is explained by the GP model.
The results of this procedure are shown in Figure~\ref{fig:cross_validation_manifold}.

Changing the number of neighbors for the Isomap algorithm has very little effect on the
recovered embedding for values between $\sim$5 and 50 neighbors. For less than 5 neighbors, we find that we
are unable to recover the third component. This is likely due to the fact that using a small number
of neighbors adds noise to the distance matrix. Removing the uncertainty cuts described in Section~\ref{sec:isomap_sample} has almost no
effect on the recovered embedding. Similarly, randomly discarding 20\% or 50\% of the SNe~Ia in the
sample does not noticeably affect the embedding.

Our sample was selected to be very high signal-to-noise, so resampling the uncertainties
of each spectrum has little effect on the recovered embedding. We find that we are able to degrade
the spectrum down to a signal-to-noise of $\sim$5 per 1000~km/s before we start to see a degradation
in the recovered embedding. Interestingly, we are able to bin the spectrum down to $\sim$10000~km/s
before we see significant changes in the embedding, which corresponds to a very low resolution
spectrum ($R\sim30$).

Finally, we investigated how changing the wavelength range of our spectra from the baseline
of 3300--8600~\AA~affects the recovered spectrum. Decreasing the cutoff wavelength at the red end
has almost no effect on the embedding, but increasing the cutoff at the blue end does, with a significant
degradation in the third component for cutoffs above $\sim$4000~\AA. As discussed in detail in
\citet{nordin18}, there is a large amount of
variability in the U-band, particularly near the \ion{Ca}{2} H\&K lines, and not having observations
of this region of the spectrum significantly degrades the performance of our model.

\section{Exploring the Intrinsic Diversity of SNe~Ia} \label{sec:discussion}

\subsection{Components of the Twins Embedding}

\begin{figure*}
\plotone{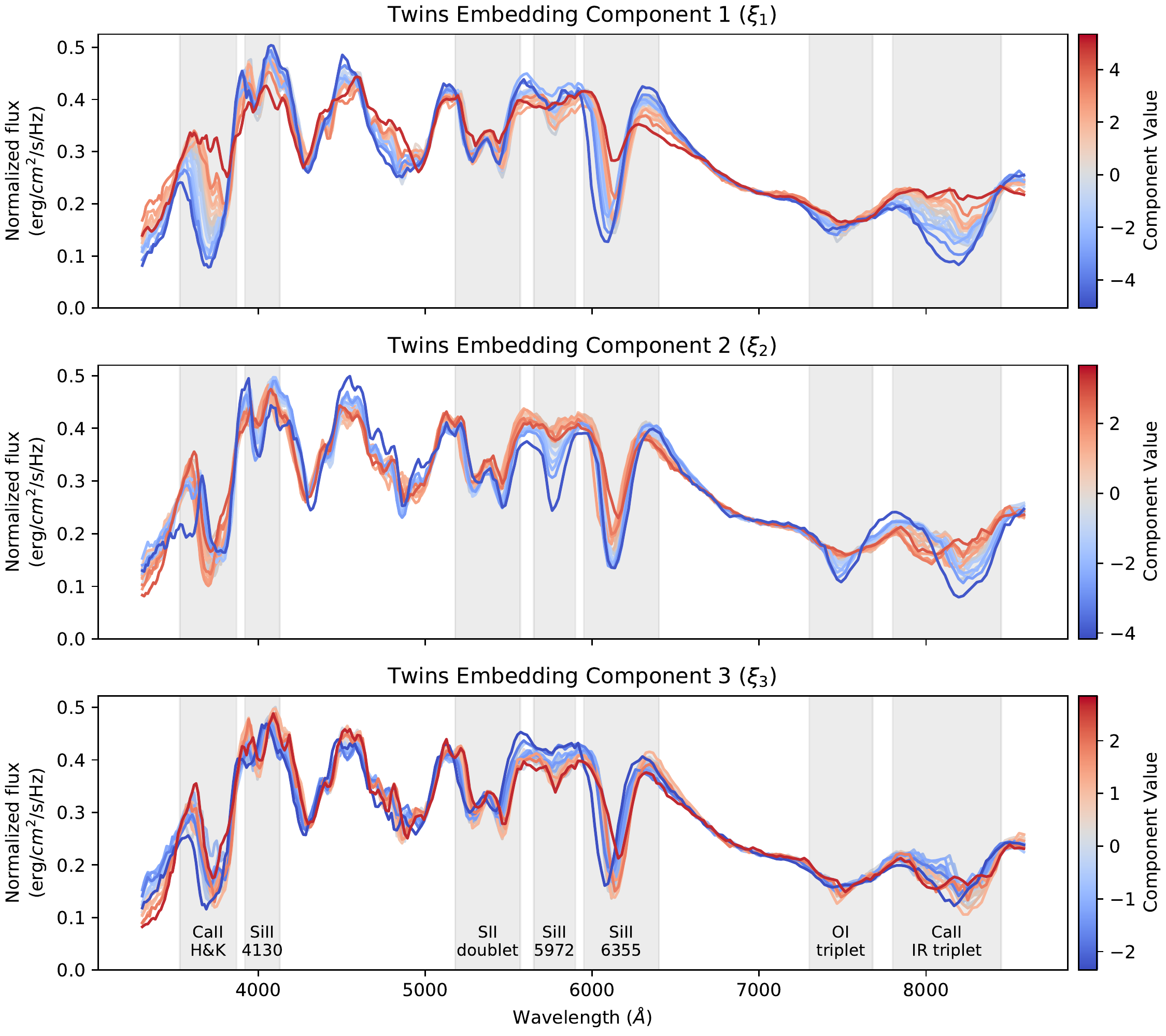}
\caption{
    Effect of each of the components of the Twins Embedding on the spectra of SNe~Ia at maximum light.
    For each component, we show the median spectrum in twenty evenly-populated bins of the coordinates of
    that component. The spectra are colored according to their component value.
}
\label{fig:component_steps}
\end{figure*}

We now examine what effects each of the three components of the Twins Embedding have on the spectrum of a SN~Ia.
For each component, we calculate the median spectrum in twenty evenly-populated bins of the coordinates
for that component.
We plot these median spectra in Figure~\ref{fig:component_steps}. We find that each component captures a different smooth,
nonlinear spectral sequence.

The first component of the Twins Embedding ($\xi_1$) primarily affects the pseudo-equivalent widths of the \ion{Ca}{2} features.
This component maps out a
full spectral sequence of the changing pseudo-equivalent width of the \ion{Ca}{2}\ H\&K feature, and a similar spectral
sequence is seen for the \ion{Ca}{2} IR triplet. Additionally, $\xi_1$ shows a spectral sequence for the \ion{Si}{2}
6355~\AA\ feature. This component is the only one that has a major impact on the emission profile of this feature,
and appears to be capturing diversity in both the optical depth and line velocity of the \ion{Si}{2}~6355~\AA\ feature.
$\xi_1$ has very little effect on the
depth of the \ion{Si}{2}~5972~\AA\ feature that is typically associated with the width of the light curve.

The second component ($\xi_2$) primarily affects the pseudo-equivalent widths of the \ion{Si}{2} lines. A spectral sequence is
identified in the \ion{Si}{2} 4130~\AA, \ion{Si}{2}~5972~\AA, and 6355~\AA\ features.
Interestingly, this component directly changes the depths of these lines without having a major effect on the
line velocities themselves. $\xi_2$ also has a large effect around 3700~\AA\ that appears to be due to a set of \ion{Si}{2}
lines at rest frame wavelengths between 3853\ and 3863~\AA. This component does not appear to affect the \ion{Ca}{2} H\&K absorption
typically associated with these wavelengths, and by comparing with $\xi_1$, we can see that the Twins Embedding
cleanly separates two modes of variability in this wavelength range. $\xi_2$ has a very complex spectral sequence
for redder wavelengths, spanning both the \ion{O}{1} and \ion{Ca}{2} features.

The third component $\xi_3$ primarily affects the ejecta velocity profiles of the supernova. This component identifies a spectral
sequence in the velocity of the \ion{S}{2} doublet feature near 5400~\AA\ and in the velocity profile of the
\ion{Si}{2}~6355~\AA\ feature. Interestingly, $\xi_3$ seems to capture an overall shift in velocity in the emission profile:
as this component is varied, we see the full absorption profile shift in wavelength for both the \ion{S}{2} doublet and the
\ion{Si}{2}~6355~\AA\ feature. This contrasts with $\xi_1$ which also alters the observed line velocity of
these emission features by increasing the optical depth of these elements at larger velocities. Similar effects are seen
for the velocities of many other lines in the spectra. $\xi_3$
does not have a major effect on any of the line depths in the spectrum. Interestingly, we see a sequence in the
pseudo-continuum level around the \ion{Si}{2}~5972~\AA\ feature that does not affect the pseudo-equivalent width
of this feature.

\subsection{Recovering Other Indicators of Intrinsic Diversity} \label{sec:recovering_other_indicators}

If the Twins Embedding truly captures all of the observable intrinsic diversity of SNe~Ia, then we would expect
to be able to recover any indicator of intrinsic diversity of SNe~Ia from some transformation of the Twins
Embedding. We tested this hypothesis by training GP models to predict a wide variety of indicators
from the Twins Embedding, and then by measuring the fraction of the variance in
those indicators that is explained by the GP models.

First, we extracted the pseudo-equivalent widths (pEWs) and line velocities of the \ion{Ca}{2} H\&K feature and several of the \ion{Si}{2}
features using the methodology described in \citet{chotard11} on each of the dereddened spectra.
For SNe~Ia that were included in the U-band analysis of \citet{nordin18}, we include their spectral
indicators uNi, uTi, uSi, and uCa at maximum light, and we also include the uCa measurement pre-peak and uTi
measurement post-peak that the authors show affect standardization.
For SNe~Ia that were included in the SUGAR analysis of \citet{leget20}, we include their
SUGAR coordinates. Similarly, for SNe~Ia that were included in the SNEMO analysis of \citet{saunders18},
we include their SNEMO7 coordinates. Finally, we compare our embedding to the SALT2 $x_1$ values measured as described
in Section~\ref{sec:manifold_dataset}. As SALT2 is not able to accurately model all of the peculiar SNe~Ia that we
have included in our sample, for this comparison, we reject SNe~Ia with bad SALT2 fits where the normalized median
absolute deviation (NMAD) of the SALT2 model residuals is more than 0.12~mag or where more than 20\% of the SALT2
model residuals have an amplitude of more than 0.2~mag.

\begin{figure}
\plotone{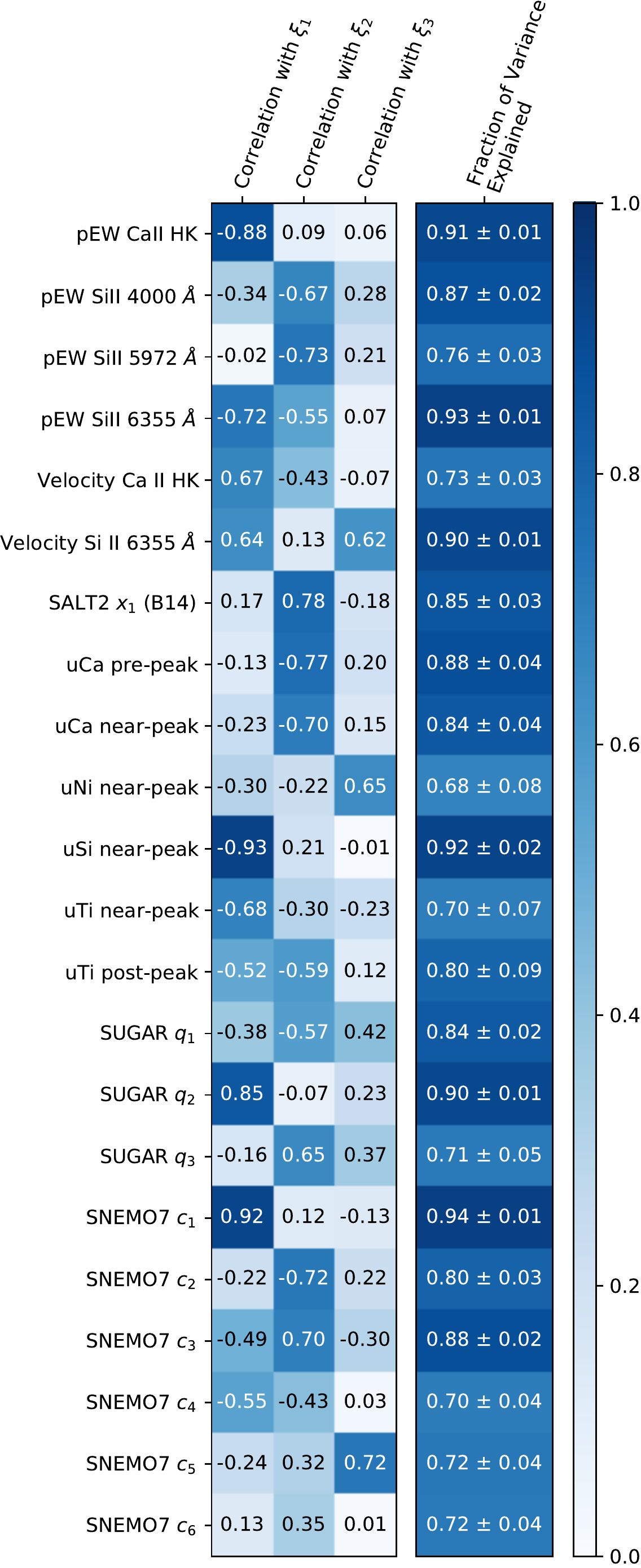}
\caption{
    Recovery of various indicators of intrinsic diversity from the Twins Embedding. In the first three columns,
    we show the Pearson correlation between each indicator and one of the components of the Twins Embedding.
    In the final column, we show the fraction of variance of each indicator that can be explained 
    using a transformation of the Twins Embedding. All of these previously established indicators
    can be recovered with high accuracy from the three-dimensional Twins Embedding.
}
\label{fig:indicators_recovery}
\end{figure}

We use the same GP model that was used to predict the flux of each spectrum over the Twins Embedding.
The details of this GP model can be found in Appendix~\ref{sec:gaussianprocess}. When available, we include
the measurement uncertainties of the indicators in the GP model.
In Figure~\ref{fig:indicators_recovery}, we show the results of this procedure. We find that the Twins Embedding
captures the majority of the diversity in all of these indicators, with $68-94$\% of the variance explained
depending on the indicator. We verified with leave-one-out GP predictions that we are able
to perform out-of-sample predictions of each of these indicators within the quoted accuracies.

Adding an additional component to the Twins Embedding does not
significantly improve the ability to recover any of these indicators while removing the third component $\xi_3$ significantly
decreases performance, further confirming that three dimensions are required to capture the diversity of SNe~Ia.
Interestingly, with transformations of the Twins Embedding, we are able to recover the majority of the variance
in each of the components of the SNEMO7 model of \citet{saunders18}. Given the location of a SN~Ia in the three-dimensional
Twins Embedding, we can therefore predict its location in the six-dimensional SNEMO7 parameter space with high accuracy.
Our results imply that the six components of SNEMO7 are not all independent and can be captured by a three-dimensional
sub-manifold. SNEMO7 is a linear model, and this is consistent with the results that we found for a simple
linear model in Section~\ref{sec:num_components}. Note that the SNEMO7 parameter space was constructed using full spectral
timeseries of SNe~Ia while the
Twins Embedding was constructed using only information at maximum light. These results then suggest that the vast
majority of of the information in the spectral timeseries of SNe~Ia is captured at maximum light.

In our comparison to the SUGAR model \citep{leget20}, we find that we are able to reproduce the first two
components of the SUGAR model with high accuracy and the majority of the variance of the third SUGAR component. Hence the
SUGAR parameter space is very similar to the Twins Embedding despite having been constructed
very differently (a linear decomposition of spectral features). We reran our algorithm to calculate the fraction of
explained variance for each indicator using a transformation of the SUGAR parameter space rather than a transformation of
the Twins Embedding. We find that both parametrizations of SNe~Ia are able to explain similar fractions of the variance
for most of the indicators that we examined. However, the SUGAR model performs significantly worse on the
U-band features of \citet{nordin18}. For example, it is only able to recover 42\% of the variance of the uNi indicator
at maximum light compared to 71\% for the Twins Embedding using the same set of SNe~Ia. This can be explained
by the fact that the SUGAR model was trained using only a specific set of spectral features, and is potentially missing
information that is not captured in those spectral features. On the other hand, the Twins Embedding was trained using
the entire spectrum at maximum light. Furthermore, the SUGAR analysis removed ``outlier SNe~Ia'' that are not
well-described by a linear model while the Twins Embedding includes all of these SNe~Ia since it is not restricted
to linear variability.

\subsection{Comparison to the Branch Classification Scheme---Diversity of Core Normal SNe~Ia}

The spectra of SNe~Ia near maximum light are often labeled in the literature using the ``Branch classification scheme''
\citep[][hereafter \citetalias{branch06}]{branch06}. In this classification scheme, the spectra of SNe~Ia are subdivided
into four classes based on the pseudo-equivalent widths (pEWs) of the \ion{Si}{2}~5972 and 6355~\AA\ absorption features.
As has been previously shown for large samples of spectra of SNe~Ia (e.g. \citet{blondin12}), the Branch classifications
do not identify distinct subtypes of SNe~Ia, and we find a continuous distribution across all of the label boundaries.

\begin{figure*}
    \epsscale{0.7}
    \plotone{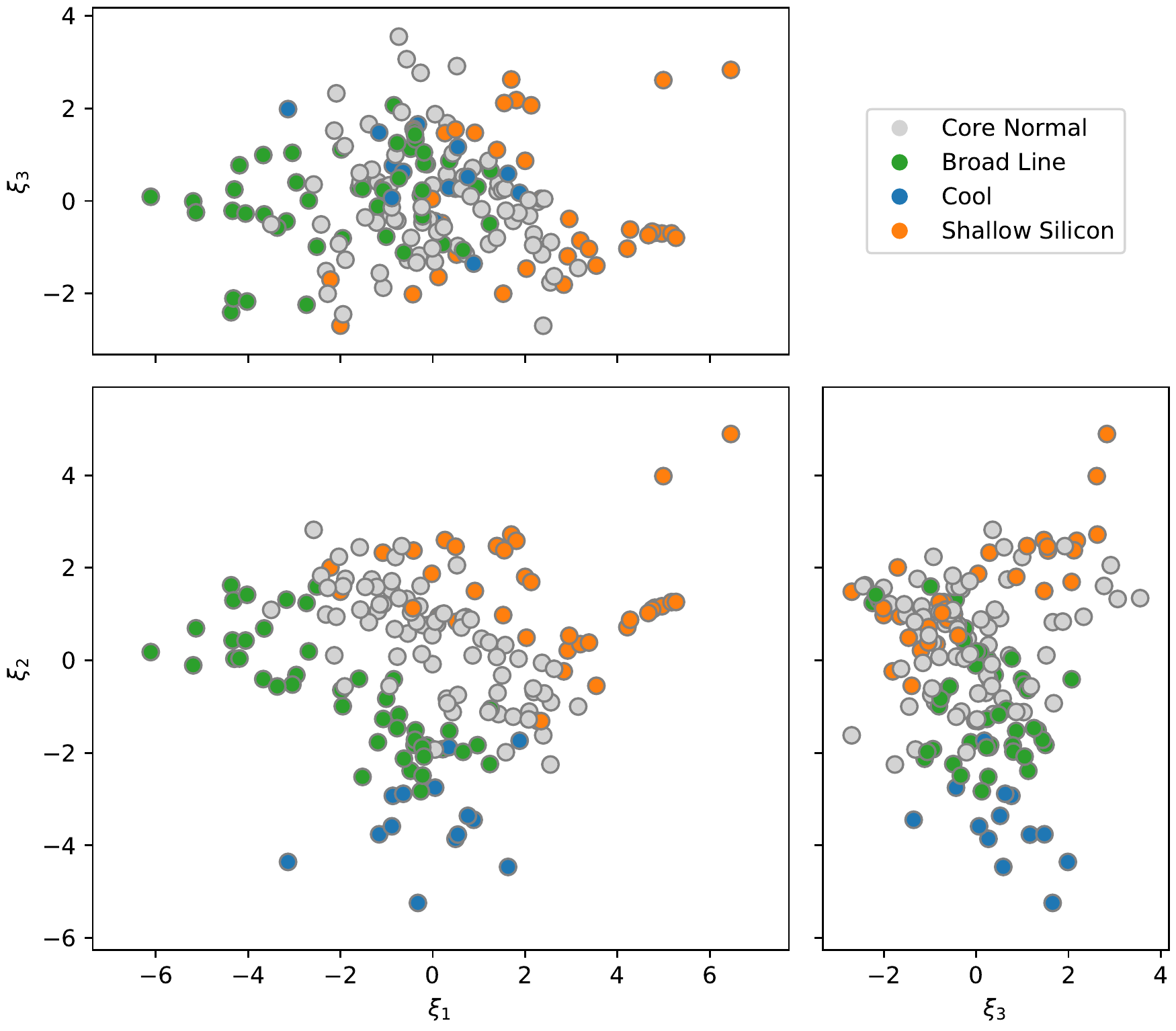}
    \caption{
        Comparison of the Branch classifications \citepalias{branch06} to the Twins Embedding.
        We find that with the first two components ($\xi_1$ and $\xi_2$) we are able to cleanly recover the Branch classification
        labels.
    }
\label{fig:branch_classification}
\end{figure*}

We compare the Branch classifications to our Twins Embedding in Figure~\ref{fig:branch_classification}.
We find that the first two components of the Twins Embedding ($\xi_1$ and $\xi_2$) cleanly separate the
different Branch classes from each
other: with a set of cuts in the Twins Embedding, we would be able to recreate the Branch classification
scheme nearly perfectly. There is some overlap at the border of each class due to
uncertainties in the pEW measurements used for the Branch classification.
Interestingly, while core normal SNe~Ia lie in a tight cluster in the parameter space used for Branch classifications,
they are spread over a fairly large region of the Twins Embedding.

\begin{figure*}
\plotone{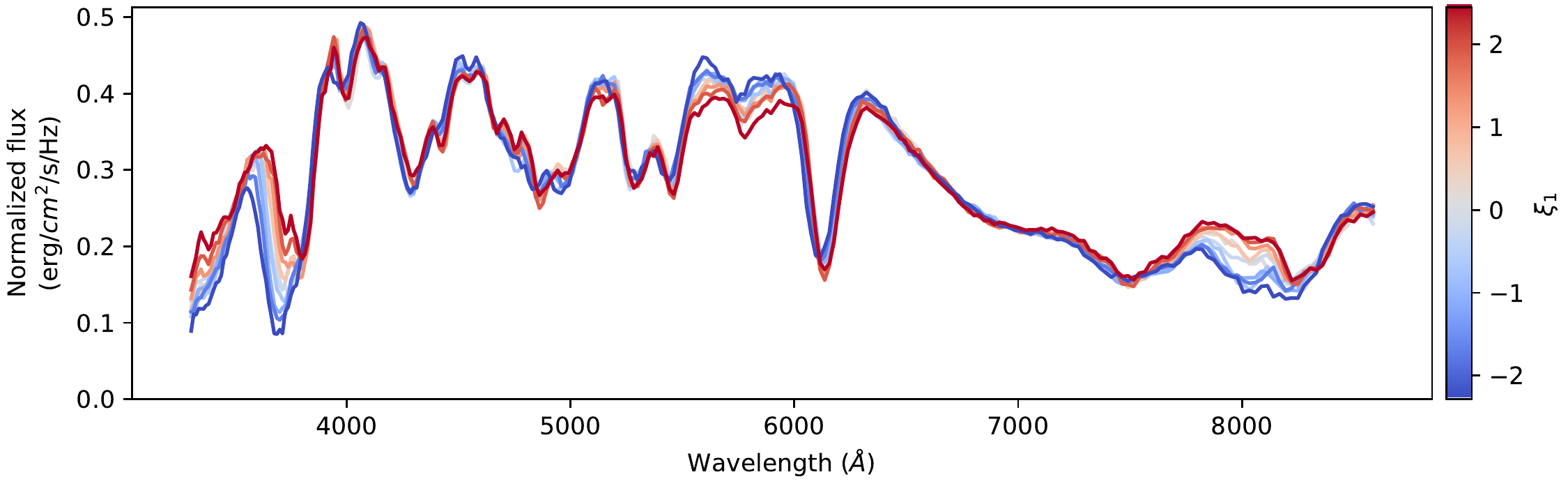}
\caption{
    Comparison of the spectra of Core Normal supernovae as a function of the first component of the Twins Embedding ($\xi_1$).
    We show the median spectrum for ten equally populated bins of $\xi_1$. We find that the Twins Embedding 
    identifies significant differences between
    Core Normal SNe~Ia. Note that there are large differences in the pseudo-continuum level near the
    \ion{Si}{2}~5972~\AA\ feature that will not be captured by analyses measuring only pseudo-equivalent widths
    or line velocities of this feature.
}
\label{fig:core_normal_comparison}
\end{figure*}

The Twins Embedding thus implies that there is significant spectral diversity among Core Normal SNe~Ia.
To probe this claim, we calculated the median spectrum for Core Normal SNe~Ia in ten equally populated bins
of $\xi_1$. The results of this procedure are shown in
Figure~\ref{fig:core_normal_comparison}. All of these median spectra have similar pEWs
for the \ion{Si}{2}~6355~\AA\ absorption feature, and have a small pEW for the \ion{Si}{2}~5972~\AA\ feature,
per the definition of a Core Normal supernova. However, the pseudo-continuum level near the
\ion{Si}{2}~5972~\AA\ feature varies quite dramatically between the different median spectra.
As analyses such as \citetalias{branch06} focused only on the pEW and ignore the pseudo-continuum level, they were
not able to identify this difference. We also see large differences for these spectra near 3800~\AA\ which we
associate with a set of \ion{Si}{2} lines between 3853 and 3863~\AA. These differences were not seen in
\citetalias{branch06} due to a lack of
spectral coverage. Finally, we see a spectral sequence for an absorption feature near 8000~\AA. This difference
was identified in \citetalias{branch06} for SN2001el, and was suggested to be due to high velocity \ion{Ca}{2}.
Hence, with only the first two components, the Twins Embedding is able to reproduce the Branch classification
scheme, and is able to identify additional diversity that is not solely limited to the pEWs and velocities of
different lines.

\subsection{Connecting Subtypes of SNe~Ia}

One major open question about SNe~Ia is whether there exist subtypes of SNe~Ia, perhaps from
different progenitor channels. To look for signs of such subtypes, we examine the distribution of the SNe~Ia in
the Twins Embedding as seen in, e.g., Figure~\ref{fig:branch_classification}. For this sample of SNe~Ia, we do not see any evidence
of large-scale bimodality in the recovered parameter space: the core of the parameter space is well-populated and
appears to be continuously filled. Note that we constructed the Twins Embedding using nonlinear dimensionality
reduction with the goal of preserving the spectral distances between ``twin'' SNe~Ia from \citet{fakhouri15}. The distances
between points in the embedding should be meaningful since we are capturing the vast majority of the variance, as shown
in Section~\ref{sec:num_components}, but the exact shape of the distribution of SNe~Ia in the embedding will depend on the algorithm
used and should not be overinterpreted.

\begin{figure*}
    \epsscale{0.7}
    \plotone{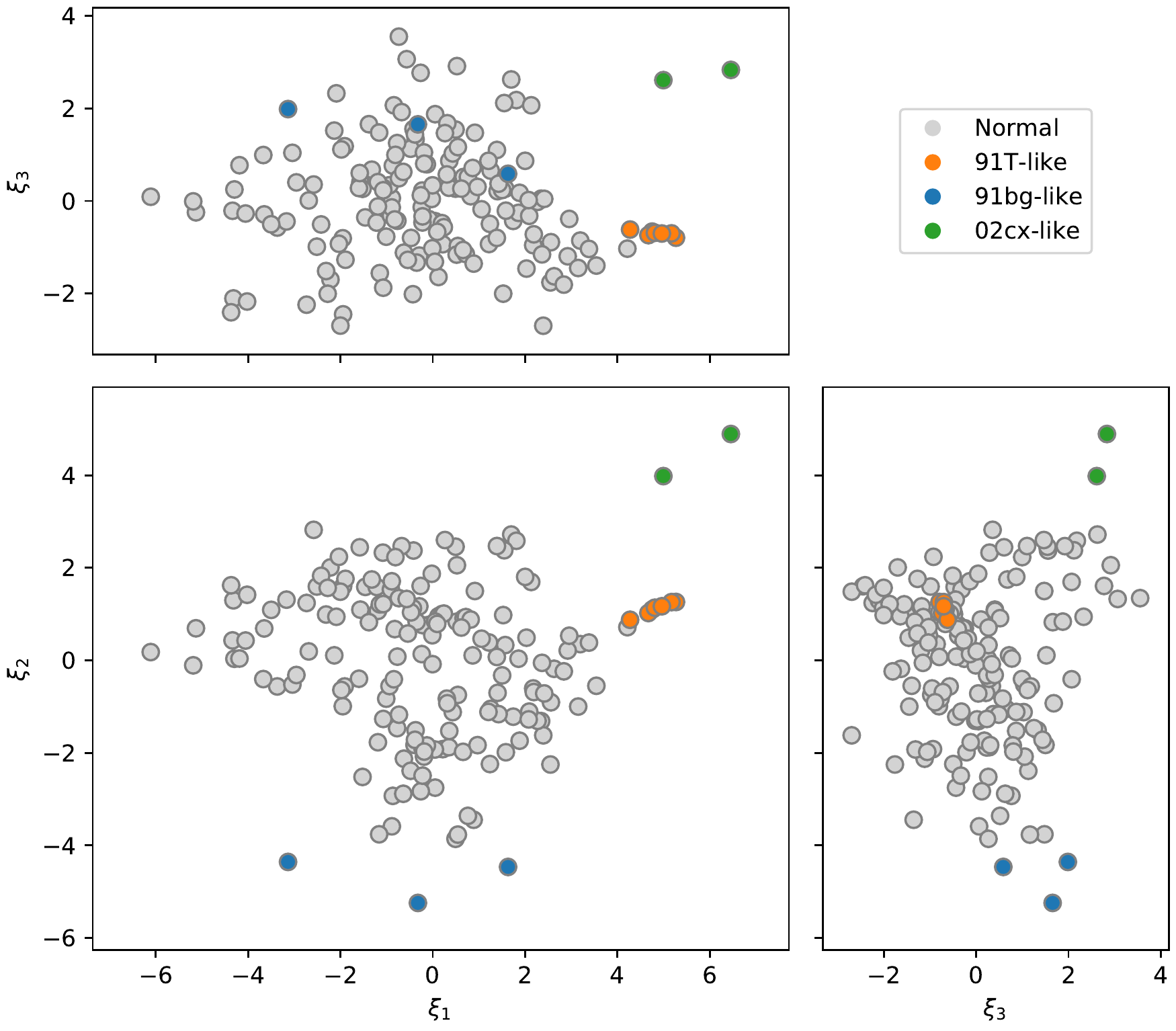}
    \caption{
        Location of peculiar SNe~Ia in the Twins Embedding.
    }
\label{fig:peculiar_locations}
\end{figure*}

The Twins Embedding includes all SNe~Ia, including ones that are typically
referred to as ``peculiar''. We identified where the peculiar subclasses of 91bg-like \citep{filippenko92a},
91T-like \citep{filippenko92b} and 02cx-like SNe~Ia \citep{li03} are located in the Twins Embedding.
We use the labeling of peculiar SNe~Ia for the SNfactory sample from \citet{lin20submitted}.
The results of this procedure can be seen in Figure~\ref{fig:peculiar_locations}.

We find that all three of these kinds of peculiar SNe~Ia lie at the edges of the Twins Embedding.
They are easily separable from
the rest of the sample using their coordinates in the Twins Embedding. There are \numtoname{\numpeculiart} 91T-like
SNe~Ia in our sample that are all tightly clustered at large values of $\xi_1$. The dereddened
spectra of the 91T-like SNe~Ia are all nearly identical which explains why they lie in such a small region
of the Twins Embedding. The \numtoname{\numpeculiarbg} 91bg-like SNe~Ia lie at very low
values of $\xi_2$ with a significant
spread in $\xi_1$. The \numtoname{\numpeculiarcx} 02cx-like SNe~Ia in our sample
are separated from the rest of the sample in all of the Twins Embedding components.

\begin{figure*}
\plotone{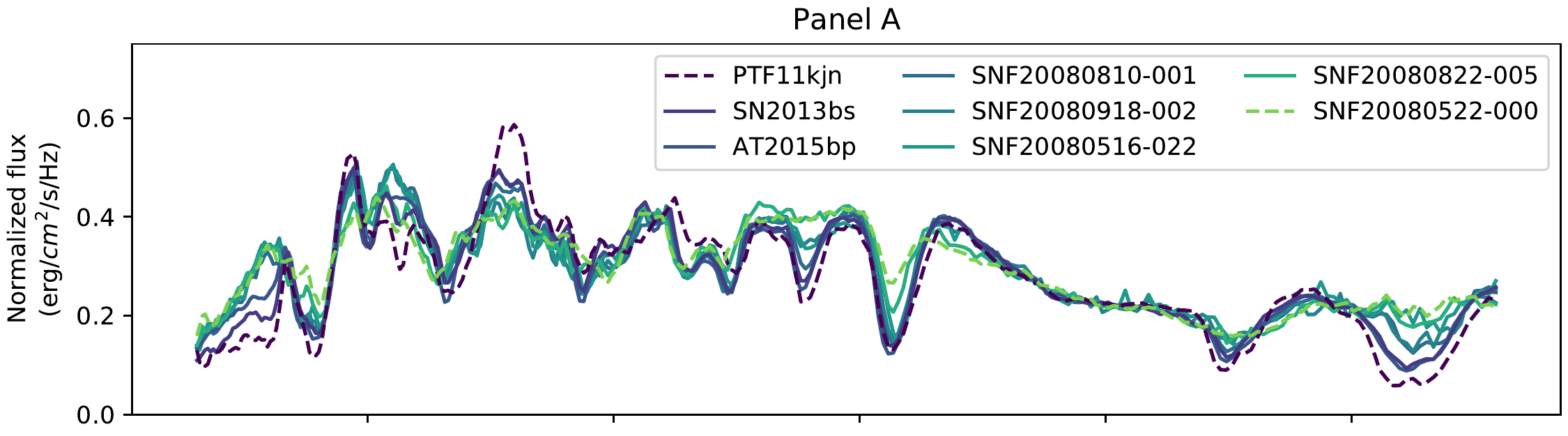}
\plotone{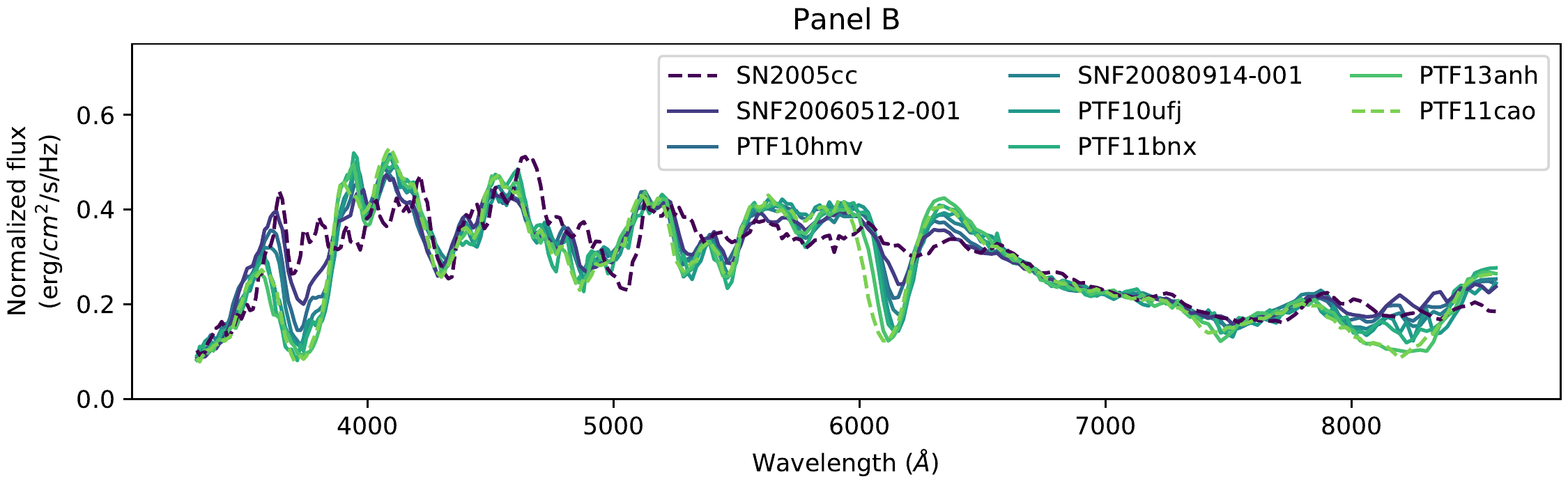}
\epsscale{0.7}
\plotone{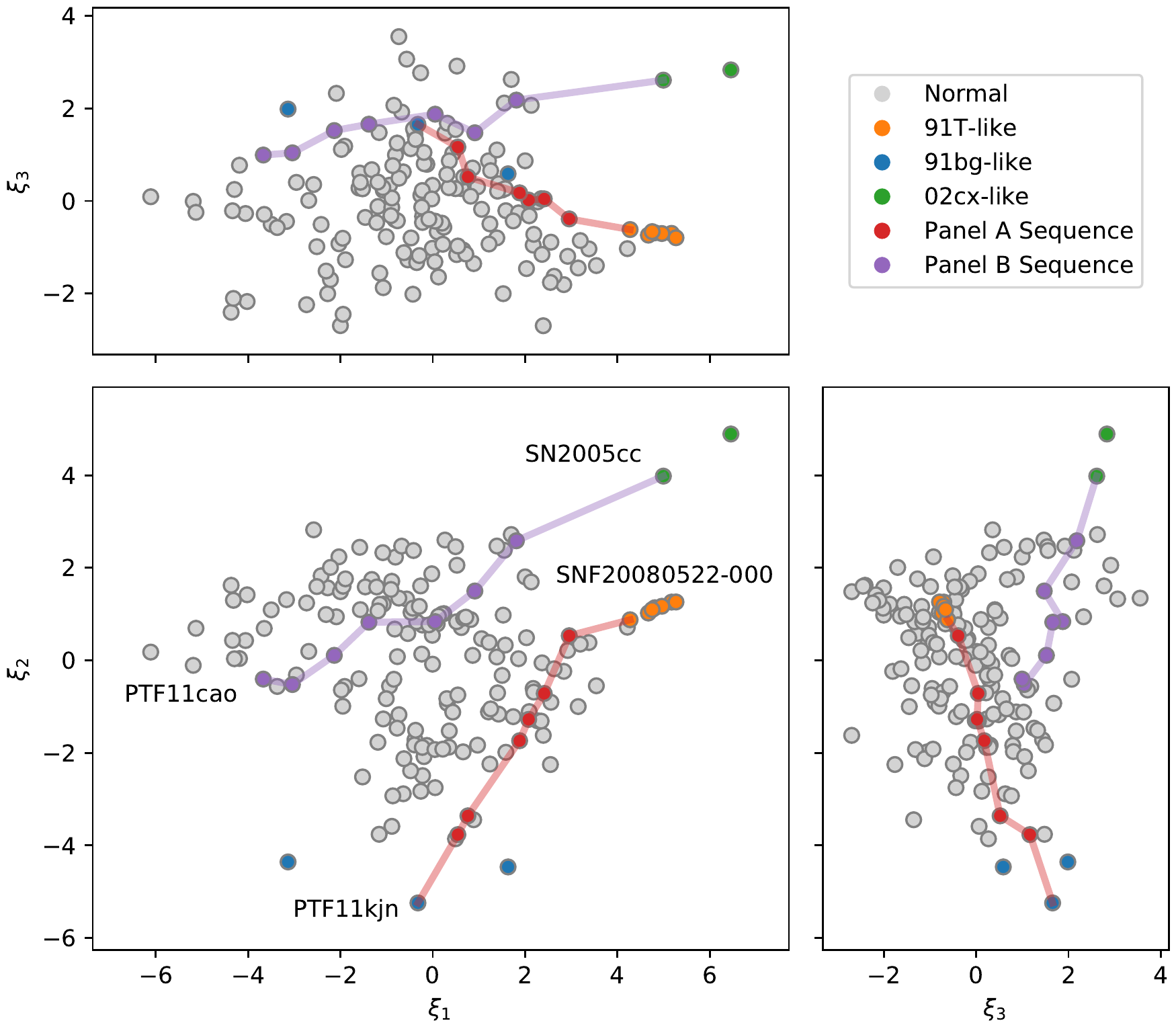}
\caption{
    Sequences of SNe~Ia near evenly-spaced coordinates in the Twins Embedding. Panel A: A sequence between
    the 91bg-like \sequencebgname\ and the 91T-like \sequencetname. These SNe~Ia map out a continuous
    spectral sequence at all wavelengths, and each pair of adjacent SNe~Ia in the sequence is a set of
    ``twins'' according to the definition of \citetalias{fakhouri15} except for \sequencebgname\ and
    \sequencebgtwin\ that are slightly above the threshold. Panel B: A sequence between the 02cx-like
    \sequencecxname\ and the ``normal'' \sequencecxend. \sequencecxname\ has large spectral differences
    relative to its closest neighbor, \sequencecxtwin, and does not follow the spectral sequence that is established at
    most wavelengths. The bottom panels show the Twins Embedding coordinates of the SNe~Ia in each of these
    sequences in red and purple respectively.
}
\label{fig:sequences}
\end{figure*}

One way to probe whether different SNe~Ia are from the same physical processes is to test
whether we can find a set of other SNe~Ia whose spectra form a continuous sequence from the first
spectrum to the second. If such a sequence can be found, then it suggests that the observed differences
between those two SNe~Ia are due to some continuous underlying physical parameter rather than being
due to discrete processes. By selecting SNe~Ia that lie along a path in the Twins Embedding, we can
identify such a sequence between any two SNe~Ia. In Panel A of Figure~\ref{fig:sequences}, we show an example of this
procedure by identifying a set of \numtoname{\sequencebgtcount} SNe~Ia that are evenly spaced between the Twins
Embedding coordinates of \sequencebgname\ (a 91bg-like SN~Ia) and \sequencetname\ (a 91T-like SN~Ia).
At each step in this sequence, the adjacent SNe~Ia are ``twins'' according to the definition of \citetalias{fakhouri15},
except for the initial pairing of \sequencebgname\ to \sequencebgtwin\ whose ``twinness percentile'' of
\sequencebgtwinpercentile\% is slightly above the threshold of 20\% defined in \citetalias{fakhouri15} due
to the low density of SNe~Ia in this region of the parameter space. This sequence of SNe~Ia is consistent at
all wavelengths: the spectral features evolve continuously when traversing the sequence of spectra,
although the changes in the spectral features are highly nonlinear.
We are able to produce similar sequences between any two 91bg-like, 91T-like or ``normal'' SNe~Ia.
This suggests that all of these SNe~Ia may come from the same underlying process, and that
their diversity could be explained by variations in some continuous underlying parameters.

On the other hand, we find that we are unable to identify continuous sequences of SNe~Ia between 02cx-like
SNe~Ia and the rest of the sample. An example of an attempt at such a sequence is shown in
Panel B of Figure~\ref{fig:sequences}. \sequencecxname\ has large spectral differences relative to the next point in the
sequence, \sequencecxtwin, and this pair falls into the \sequencecxtwinpercentile$^{th}$ percentile of \citetalias{fakhouri15}
twinness which is worse than a random matching. The spectral differences that are seen do not follow the established
sequence from the other SNe~Ia, especially between 3800 and 5000~\AA. Similar results are found for the other 02cx-like
SN~Ia in our sample, SN2011ay. This could be suggestive of a different underlying process, but it is difficult to make
definitive conclusions with our small sample of only \numtoname{\numpeculiarcx} 02cx-like SNe~Ia since we could
simply be lacking intermediate SNe~Ia in our sample.

\section{Conclusions} \label{sec:manifold_conclusions}

In this work, we showed how the diversity of the spectra of SNe~Ia at maximum light can be decomposed into its
various components. In Section~\ref{sec:maximum_estimation}, we showed how we can estimate the
spectrum at maximum light for a SN~Ia using spectra within five days of
maximum light using a model of the differential time evolution of SNe~Ia near maximum light.
This method is completely agnostic to extrinsic sources of diversity such as dust extinction or distance
uncertainties. We find that 84.6\% of the variance in the evolution of SNe~Ia near maximum
light is common to all SNe~Ia.

In Section~\ref{sec:reading_between_the_lines}, we developed a new technique to estimate extrinsic contributions
to the spectra of SNe~Ia at maximum light that we call ``Reading Between the Lines'' (RBTL). This technique
effectively uses the regions of the spectra of SNe~Ia between major spectral features
where there is little intrinsic variability to estimate the extrinsic effects on the spectrum such as the overall brightness
and reddening due to dust. We find that SNe~Ia are incredibly homogeneous in regions of the spectra between
major spectral features. For example, for the region between 6600 and 7200~\AA\ we find an intrinsic dispersion
of only $\sim$0.02~mag. After the RBTL procedure, we are left with a set of dereddened spectra of SNe~Ia at
maximum light that only have intrinsic diversity remaining.

In Section~\ref{sec:decomposing_intrinsic}, we showed how manifold learning can be used to map out the intrinsic
diversity of SNe~Ia by building a parameter space from pairs of twin SNe~Ia that we call the Twins Embedding.
We find that a three-dimensional parametrization captures \isomapgpexpvariii\% of the intrinsic diversity
of SNe~Ia at maximum light. The three components of the Twins Embedding primarily affect the \ion{Ca}{2} features,
the \ion{Si}{2} features, and the photosphere expansion velocities respectively. From this parameter space,
we are able to recover the majority of the variance in a wide range of previously-established indicators of
intrinsic diversity of SNe~Ia. We are able to reproduce the classifications of \citet{branch06} with the first two
components of the Twins Embedding, and can also identify additional diversity among Core Normal SNe~Ia that is not
captured in the classification of \citet{branch06}.

This analysis included the full range of SNe~Ia observed by the SNfactory, including
so-called ``peculiar'' SNe~Ia. We showed that we can use the Twins Embedding to identify
spectral sequences between any two SNe~Ia, including 91bg-like and 91T-like SNe~Ia but excluding 02cx-like SNe~Ia.
This suggests that the diversity of these SNe~Ia could be due to continuous variation in some underlying physical
process rather than being due to different discrete processes, although we cannot make definitive conclusions
with the limited size of our current dataset.

One limitation of our analysis is that we performed a sequential decomposition of the diversity
of spectra at maximum light rather than a simultaneous one. This will introduce additional uncertainty
into the model, but we used very high signal-to-noise observations for this analysis so each step of the model
is very well constrained. Another potential concern is that our model is not flexible enough to capture
the full diversity of SNe~Ia. For example, the phase evolution
near maximum light likely varies for SNe~Ia at different locations in the Twins Embedding. This will not
affect our analysis, as discussed in Section~\ref{sec:maximum_estimation}, because we modeled the unexplained
dispersion and found that it is very small compared to the effects that we model in the rest of the analysis.
In a future analysis, we plan on developing a hierarchical model that can simultaneously fit all
of the components of the model described here.

This analysis only considered spectra of SNe~Ia at maximum light. There could be additional intrinsic diversity
in the light curves of SNe~Ia that is not captured in the spectra at maximum light, although 
\citetalias{fakhouri15} showed that the twinning procedure performs as well using only spectra near maximum
light as it does with the full spectral timeseries and we are able to recover the majority of the variance
in the SNEMO7 components of \citet{saunders18} that were trained on the full spectral timeseries. The range
of additional intrinsic diversity at other
phases could be probed by comparing the full timeseries of SNe~Ia at similar locations in the
Twins Embedding. A similar concern is diversity in the extrinsic variability of SNe~Ia, such as variability in the dust $R_V$. As
discussed in Section~\ref{sec:rbtl_model}, the construction of the Twins Embedding is effectively insensitive
to $R_V$ variation. However, $R_V$ variation could be probed by comparing SNe~Ia at similar locations in the
Twins Embedding.

In Article~II, we show how the Twins Embedding can be used to improve standardization of SNe~Ia.
To summarize its conclusions, there is significant diversity in the luminosity of SNe~Ia that is
not explained by light curve width and color but that can be identified using the Twins Embedding. Taking
this additional information about intrinsic diversity into account reduces uncertainties in the distance
estimates to SNe~Ia and reduces correlations with host galaxy properties.

All of the code used in this analysis is publicly available at
\url{https://doi.org/10.5281/zenodo.4670772}, and the data are available
on the SNfactory website at \url{https://snfactory.lbl.gov/snf/data/}.

\section{Acknowledgements}

We thank the technical staff of the University of Hawaii 2.2-m telescope, and Dan Birchall for observing assistance.
We recognize the significant cultural role of Mauna Kea within the indigenous Hawaiian community, and we appreciate
the opportunity to conduct observations from this revered site.
This work was supported in part by the Director, Office of Science, Office of High Energy Physics of the U.S.
Department of Energy under Contract No. DE-AC025CH11231.
Additional support was provided by NASA under the Astrophysics Data
1095 Analysis Program grant 15-ADAP15-0256 (PI: Aldering).
Support in France was provided by CNRS/IN2P3, CNRS/INSU, and PNC; LPNHE acknowledges support from LABEX ILP,
supported by French state funds managed by the ANR within the Investissements d’Avenir programme under
reference ANR-11-IDEX-0004-02.
Support in Germany was provided by DFG through
TRR33 “The Dark Universe” and by DLR through grants FKZ 50OR1503
and FKZ 50OR1602.
In China support was provided by Tsinghua University
985 grant and NSFC grant No. 11173017.
We thank the Gordon and Betty Moore Foundation for
their continuing support.
This project has received funding from the European Research Council (ERC)
under the European Union’s Horizon 2020 research and innovation programme
(grant agreement No. 759194 – USNAC).

\appendix

\section{Gaussian Process Regression} \label{sec:gaussianprocess}

We use Gaussian process (GP) regression to model how different properties of SNe~Ia
vary across the Twins Embedding. A GP is stochastic process over some set of continuous
inputs where observations at any finite set of inputs follow a multivariate normal
distribution. For a comprehensive discussion of GPs, see \citet{rasmussen06}. In our case,
we use GPs to produce nonparametric models of how different properties of SNe~Ia vary
over the Twins Embedding.

We assume that the covariance between observations $f_i$ and $f_j$ at respective coordinates
$\vec{x}_i$ and $\vec{x}_j$ can be described using the following function referred to as a ``kernel'':

\begin{align}
    K\left(\vec{x}_i, \vec{x}_j; A, l, \eta\right) = A^2 \left(1 + \sqrt{3 \frac{\lVert \vec{x}_i - \vec{x}_j \rVert^2}{l^2}}\right) \exp\left(-\sqrt{3 \frac{\lVert \vec{x}_i - \vec{x}_j \rVert^2}{l^2}}\right) + \eta^2 \delta_{ij}
\end{align}

The first term of this equation is a Mat\'ern 3/2 kernel that captures the covariance between
observations at nearby coordinates with hyperparameters for the amplitude $A$ and the length scale $l$.
The second term captures uncorrelated variance for repeated observations at the same coordinates with
the intrinsic dispersion represented by the hyperparameter $\eta$.

We include an additional hyperparameter $\mu$ in our GP model representing a constant
mean value. A set of observations $\vec{f}$ are then described by:

\begin{align}
    \vec{f} \sim \mathcal{N}(\mu, K(\vec{x}; A, l, \eta))
\end{align}

For each property, we optimize the hyperparameters of the model to maximize the likelihood.
We implement this model using the \texttt{George} package \citep{ambikasaran15}. The final model
can then be used to estimate the value of the property at any location in the Twins Embedding by
conditioning the GP on the observations of that property.

\bibliography{references}{}
\bibliographystyle{aasjournal}

\end{document}

%% file: latex/attrition_variables.tex
\newcommand{\nummanifoldsne}{203}
\newcommand{\nummanifoldspectra}{598}
\newcommand{\numinterpsne}{173}

%% file: latex/peculiar_counts.tex
\newcommand{\numpeculiarbg}{3}
\newcommand{\numpeculiart}{7}
\newcommand{\numpeculiarcx}{2}

%% file: latex/sequence_peculiars.tex
\newcommand{\sequencebgname}{PTF11kjn}

\newcommand{\sequencebgtwin}{SN2013bs}
\newcommand{\sequencebgtwinpercentile}{29}
\newcommand{\sequencebgtcount}{8}
\newcommand{\sequencetname}{SNF20080522-000}

%% file: latex/sequence_peculiars_cx.tex
\newcommand{\sequencecxname}{SN2005cc}

\newcommand{\sequencecxtwin}{SNF20060512-001}
\newcommand{\sequencecxtwinpercentile}{64}
\newcommand{\sequencecxend}{PTF11cao}

%% file: latex/explained_variance.tex
\newcommand{\isomapgpexpvarfulliii}{86.6}

\newcommand{\isomapgpexpvariii}{89.2}

\newcommand{\isomapgpexpvarindivi}{49.3}
\newcommand{\isomapgpexpvarindivii}{26.2}
\newcommand{\isomapgpexpvarindiviii}{11.1}

\newcommand{\isomapgpnoise}{2.9}

%% file: latex/attrition_table.tex
\textbf{General selection requirements} & \\
Initial sample (SNe Ia with at least 5 SNfactory spectra) & 280 \\
SALT2 date of maximum light uncertainty < 1 day           & 272 \\
At least one spectrum within 5 days of maximum light      & 250 \\
At least one spectrum with S/N 3300-3800~\AA\ > 100       & 203 \\
\hline
\textbf{Manifold learning selection requirements} & \\
\textbf{(Section~\ref{sec:isomap_sample})} & \\
Spectrum at max. uncertainty < 10\% of intrinsic variance        & 173 \\

%% file: latex/differential_time_evolution_parameters_short.tex
3463 & 0.102 & 0.01134 & 0.061 & 0.024 & 0.062 & 0.083 & 0.160 \\
3801 & 0.049 & 0.00577 & 0.069 & 0.023 & 0.035 & 0.095 & 0.246 \\
4171 & -0.001 & 0.00944 & 0.035 & 0.011 & 0.059 & 0.120 & 0.072 \\
4579 & -0.004 & 0.00695 & 0.037 & 0.009 & 0.027 & 0.061 & 0.066 \\
5026 & 0.014 & 0.00683 & 0.047 & 0.024 & 0.041 & 0.052 & 0.049 \\
5516 & -0.003 & 0.00601 & 0.055 & 0.008 & 0.044 & 0.056 & 0.089 \\
6055 & -0.024 & 0.00686 & 0.092 & 0.042 & 0.033 & 0.084 & 0.252 \\
6645 & -0.011 & 0.00337 & 0.062 & 0.026 & 0.007 & 0.017 & 0.019 \\
7294 & 0.020 & -0.00046 & 0.075 & 0.046 & 0.022 & 0.039 & 0.047 \\
8006 & 0.003 & 0.00584 & 0.098 & 0.054 & 0.022 & 0.069 & 0.191 \\

%% file: latex/twins_manifold_coordinates_short.tex
SNF20050728-006 & 0.46 & 0.31 & 0.117 & 0.035 & 0.276 & 0.014 & -0.094 & 0.039 & -0.937 & -0.557 & 0.339 \\
SNF20050729-002 & 0.92 & 0.34 & -0.107 & 0.038 & -0.260 & 0.018 & 0.128 & 0.030 & 0.621 & -0.536 & 0.152 \\
SNF20050821-007 & 0.42 & 0.29 & -0.043 & 0.036 & -0.205 & 0.014 & 0.029 & 0.040 & -2.580 & 2.821 & 0.355 \\
SNF20050927-005 & 0.19 & 0.35 & 0.018 & 0.035 & 0.033 & 0.014 & 0.301 & 0.056 & -0.176 & -1.688 & 1.221 \\
SNF20051003-004 & 1.16 & 0.16 & -0.100 & 0.030 & -0.349 & 0.013 & 0.071 & 0.065 & 2.030 & 0.489 & -1.463 \\
SNF20060511-014 & -0.78 & 0.18 & -0.041 & 0.035 & -0.030 & 0.013 & 0.037 & 0.048 & -0.396 & -1.828 & 1.501 \\
SNF20060512-001 & 0.40 & 0.14 & -0.007 & 0.030 & 0.046 & 0.012 & -0.096 & 0.057 & 1.810 & 2.584 & 2.179 \\
SNF20060512-002 & -0.26 & 0.23 & 0.086 & 0.033 & 0.079 & 0.015 & -0.197 & 0.045 & -1.993 & 1.080 & -1.411 \\
SNF20060521-001 & -1.63 & 0.29 & -0.111 & 0.036 & -0.057 & 0.013 & 0.047 & 0.034 & 1.238 & -2.242 & -0.497 \\
SNF20060521-008 & -1.35 & 0.32 & 0.122 & 0.042 & 0.404 & 0.016 & -0.206 & 0.041 & -3.916 & -0.466 & -1.604 \\